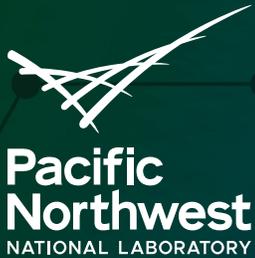

# DECODING THE MOLECULAR UNIVERSE

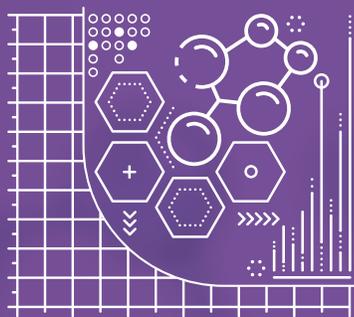

*m/q* INITIATIVE
@PNNL

## 2023 WORKSHOP

# PREFACE

On August 9-10, 2023, a workshop was convened at the Pacific Northwest National Laboratory (PNNL) in Richland, WA that brought together a group of internationally recognized experts in metabolomics, natural products discovery, chemical ecology, chemical and biological threat assessment, cheminformatics, computational chemistry, cloud computing, artificial intelligence, and novel technology development.

These experts were invited to assess the value and feasibility of a grand-scale project to create new technologies that would allow the identification and quantification of all small molecules, or to "decode the molecular universe." The Decoding the Molecular Universe project would extend and complement the success of the Human Genome Project (HGP) by developing new capabilities and technologies to measure small molecules – defined as non-protein, non-polymer molecules less than ~1500 Daltons – of any origin and generated in biological systems or produced abiotically.

The workshop was inspired by recognition that these new technologies would drive the next revolution in multiple fields of strategic importance to the nation: biological sciences, medical sciences, national security, chemistry, and the bio-economy. The workshop was supported by PNNL's investment to advance the field of computational mass spectrometry in support of Department of Energy and Department of Homeland Security missions via a project named the *m/q* Initiative. The workshop was modeled on similar workshops and related efforts held in the mid-1980s that were the catalyst for the HGP.

Workshop participants and observers included representatives from academia, government agencies, industry, and national laboratories whose research involves or where the success of programs and mission areas depends heavily on the study or measurement of such small molecules.

Primary motivations for the workshop were a nearly universal need – across scientific disciplines and applied and basic research domains – to identify millions of small molecules that collectively hold the key to understanding systems of national importance, as well as a 20-year history of mostly incremental advancements in measurement and chemical identification technologies – a situation surprisingly like that for DNA sequencing prior to the HGP. Despite the apparent diversity in instrumentation used for performing such measurements, available technologies share common challenges in determining which small molecules may be present and in what quantities. As a consequence, how the millions of small molecules present in the world contribute to human disease or illness, and the functions of the increasing number of biological systems creating our next generation of biofuels, bioproducts, medicines and crops, remain a mystery.

In contrast, technologies for broadly measuring proteins in high throughput – i.e., proteomics – are much more mature, in large part due to the high correlation of the proteome to the genome:  proteins are direct read outs of the genetic code, and if the genome is known, then the proteome can be predicted; proteins have minimal chemical complexity, consisting of 20 amino acids; and proteins, as their constituent peptides, behave characteristically and predictably during analysis by mass spectrometry.

The ultimate goal of the workshop was to establish a founding coalition of experts that would continue to define, refine and launch a grand scale Decoding the Molecular Universe project.

To assess the need, value, and feasibility of decoding the molecular universe, workshop attendees:

1) explored what new understanding of biological and environmental systems could be revealed through the lens of small molecules;

2) characterized the similarities in current needs and technical challenges between each science or mission area (biomedical, environmental, chemical/biological defense, etc.) for unambiguous and comprehensive determination of the composition and quantities of small molecules of any sample;

3) determined the extent to which technologies or methods currently exist for unambiguously and comprehensively determining the small molecule composition (quantitatively and qualitatively) of any sample and in a reasonable time; and

4) identified the attributes of the ideal technology or approach for universal small molecule measurement and identification.

The workshop concluded with a discussion of how a project of this scale could be undertaken, possible thrusts for the project, early proof-of-principle applications, and similar efforts upon which the project could be modeled.

# Decoding the Molecular Universe – Workshop Report

August 9-10, 2023

Chair: Thomas O. Metz, Biological Sciences Division, Earth and Biological Sciences Directorate, Pacific Northwest National Laboratory, Richland, WA, USA

Co-Chair: Robert G. Ewing, Nuclear, Chemistry and Biology Division, National Security Directorate, Pacific Northwest National Laboratory, Richland, WA, USA

**Prepared by:**

| | |
|---|---|
| Thomas O. Metz | Biological Sciences Division, Earth and Biological Sciences Directorate, Pacific Northwest National Laboratory, Richland, WA, USA |

**With contributions from:**

| | |
|---|---|
| Joshua N. Adkins | Biological Sciences Division, Earth and Biological Sciences Directorate, Pacific Northwest National Laboratory, Richland, WA, USA |
| Peter B. Armentrout | Department of Chemistry, University of Utah, Salt Lake City, UT, USA |
| Patrick Chain | Biosciences Division, Chemical Earth and Life Sciences Directorate, Los Alamos National Laboratory, Los Alamos, NM USA |
| Fanny Chu | Nuclear, Chemistry and Biology Division, National Security Directorate, Pacific Northwest National Laboratory, Richland, WA, USA |
| Courtney D Corley | AI and Data Analytics Division, Pacific Northwest National Laboratory, Richland, WA, USA |
| John R. Cort | Biological Sciences Division, Earth and Biological Sciences Directorate, Pacific Northwest National Laboratory, Richland, WA, USA |
| Elizabeth Denis | Nuclear, Chemistry and Biology Division, National Security Directorate, Pacific Northwest National Laboratory, Richland, WA, USA |
| Daniel Drell (retired) | Office of Biological and Environmental Research, U.S. Department of Energy, Washington, DC |
| Katherine R. Duncan | University of Strathclyde, Strathclyde Institute of Pharmacy and Biomedical Sciences, Glasgow, UK |
| Robert G. Ewing | Nuclear, Chemistry and Biology Division, National Security Directorate, Pacific Northwest National Laboratory, Richland, WA, USA |
| Facundo M. Fernandez | School of Chemistry and Biochemistry and Petit Institute for Bioengineering and Bioscience, Georgia Institute of Technology, Atlanta, GA, USA |
| Oliver Fiehn | University of California Davis, West Coast Metabolomics Center, Davis, CA, USA |
| Neha Garg | School of Chemistry and Biochemistry and Petit Institute for Bioengineering and Bioscience, Center for Microbial Dynamics and Infection, Georgia Institute of Technology, Atlanta, GA, USA |
| Stefan Grimme | Mulliken Center for Theoretical Chemistry, Clausius Institute for Physical and Theoretical Chemistry, University of Bonn, Germany |
| Christopher Henry | Data Science and Learning Division, Argonne National Laboratory, Lemont, IL, USA |
| Robert L. Hettich | Biosciences Division, Oak Ridge National Laboratory, Oak Ridge, TN, USA |
| Tobias Kind | Enveda Biosciences, Boulder, CO, USA |
| Roger G. Linington | Department of Chemistry, Simon Fraser University, Burnaby, BC, Canada |
| Gary W. Miller | Department of Environmental Health Sciences, Mailman School of Public Health, Columbia University, New York, NY |
| Trent Northen | Environmental Genomics and System Biology Division, Lawrence Berkeley National Laboratory, CA, USA |
| Kirsten Overdahl | Metabolomics Core Facility, National Institute of Environmental Health Sciences, Research Triangle Park, NC, USA |
| Ari Patrinos | Novim Group, Santa Barbara, CA, USA |
| Daniel Raftery | Northwest Metabolomics Research Center, Department of Anesthesiology and Pain Medicine, University of Washington, Seattle, WA, USA |
| Paul Rigor | Center for Cloud Computing, Pacific Northwest National Laboratory, Richland, WA, USA |
| Richard D. Smith | Biological Sciences Division, Earth and Biological Sciences Directorate, Pacific Northwest National Laboratory, Richland, WA, USA |
| Jon Sobus | Advanced Analytical Chemistry Methods Branch, Chemical Characterization and Exposure Division, Center for Computational Toxicology and Exposure, United States Environmental Protection Agency, Research Triangle Park, NC, USA |
| Justin Teeguarden | Environmental Molecular Sciences Laboratory, Pacific Northwest National Laboratory, Richland, WA, USA |
| Akos Vertes | Department of Chemistry, George Washington University, Washington, DC, USA |
| Katrina Waters | Biological Sciences Division, Earth and Biological Sciences Directorate, Pacific Northwest National Laboratory, Richland, WA, USA |
| Bobbie-Jo Webb-Robertson | Biological Sciences Division, Earth and Biological Sciences Directorate, Pacific Northwest National Laboratory, Richland, WA, USA |
| Antony Williams | Computational Chemistry & Cheminformatics Branch, Chemical Characterization and Exposure Division, Center for Computational Toxicology and Exposure, United States Environmental Protection Agency, Research Triangle Park, NC, USA |
| David Wishart | Department of Biological Sciences, University of Alberta, Edmonton, AB, Canada |

**INTRODUCTION**

In the mid-1980s, several events occurred that led to the conceptualization and eventual initiation of the world's largest collaborative biological project – the Human Genome Project (HGP). In 1985, Robert Sinsheimer – then Chancellor of the University of California, Santa Cruz – convened a workshop with 18 participants to discuss the feasibility of sequencing the human genome, and then wrote and distributed a workshop summary.[1] Also in 1985, Charles DeLisi, who had just left a senior investigator position at the NIH to become the director of the Health and Environmental Research Programs at the U.S. Department of Energy, asked Mark Bitensky of Los Alamos National Laboratory to organize a workshop to discuss the same topic, which took place March 3-4, 1986 in Santa Fe.[2] Later that year, Renato Dulbecco published a perspective in the journal Science on the knowledge that a fully sequenced human genome would provide in the fight against cancer.[3]

Other similar events followed, and extensive efforts occurred behind the scenes – all of which culminated in a $13-million line item in President Ronald Reagan's 1987 budget request that was realized in 1988 as the first official funding of the HGP. After 13 years and $3 billion, the HGP resulted in a 90% completion of the human genome sequence and innumerable increases in understanding of the roles of genes in human biology and environmental systems (e.g., soil microbial communities), key advancements in gene and transcript sequencing capabilities, and substantial beneficial economic impacts.[4] Yet an individual's genetics alone has been shown to explain just 40% of disease,[5] with environmental exposures[6] (e.g., via chemicals) and interactions of such with the genome assumed to explain the remainder.[7]

In contrast, the single most-focused U.S. effort to date towards establishing capabilities that would enable complete knowledge of the role of small molecules – defined as non-protein, non-polymer molecules less than ~1500 Daltons – in biological systems was the NIH Common Fund Metabolomics Program, a 10-year (2013-2023) and ~$200-million effort focused on establishing a national capacity in and advanced capabilities for metabolomics in biomedical applications and which was roughly 6.7% of the investment made towards sequencing the human genome. Many groundbreaking and significant resources (e.g., the National Metabolomics Data Repository - the Metabolomics Workbench[8]) and technological advancements were made and new biological understanding achieved under that program, particularly in the understanding of the role of metabolism and metabolites in cancer.[9-14]

However, the scientific community is still unable to ascertain the complete chemical composition and quantities of small molecules in any given sample, which is one of the last remaining steps towards achieving a complete and foundational understanding of biological and environmental systems. This is likely because the fundamental analytical paradigms implemented in high throughput measurements of small molecules (e.g., 'metabolomics') have not changed in over 50 years.

The first attempts to perform modern high throughput analysis of small molecules in biological systems were made by Linus Pauling and Arthur Robinson in the late 1960s and early 1970s in support of Pauling's concept of 'orthomolecular medicine.'[15,16] Pauling and Robinson utilized chromatography coupled with flame ionization detection and then mass spectrometry to detect and quantify the small molecule components of biofluids such as urine, and then performed pattern recognition analysis to identify signatures that were correlated to a patient's phenotype. Many significant advancements in measuring small molecules have been made since,[17] and such measurements are now performed with orders of magnitude higher throughput, higher quality (e.g., in terms of accuracy and precision), and more molecules routinely measured[18-20]

(**Figure 1**). However, the general analytical approach has not dramatically changed – samples are still extracted, profiled using either chromatography coupled with mass spectrometry or NMR, molecules are identified by matching experimental data to reference data obtained through analysis of chemical standards, and data are frequently interpreted initially via pattern recognition approaches.

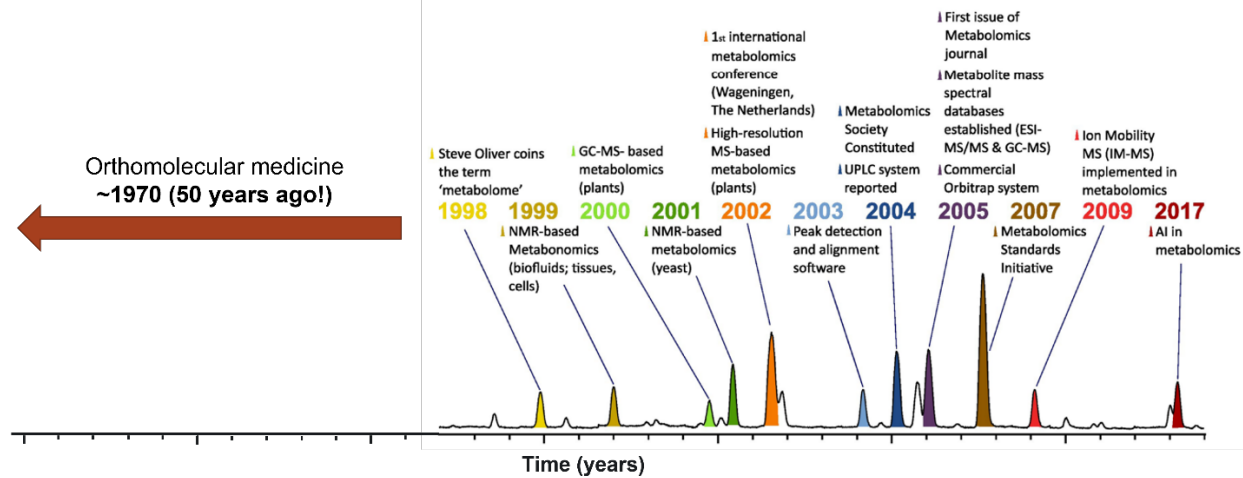

**Figure 1. Timeline of major conceptual, technical, and community advancements in metabolomics.** Adapted with permission from the authors from Reference 17 under the Creative Commons Attribution-NonCommercial-NoDerivatives License 4.0 (CC BY-NC-ND).

The Decoding the Molecular Universe workshop was convened to critically examine the current analytical state-of-the-art implemented in metabolomics and other small molecule analyses, speculate on what new understandings of biological and environmental systems might be gained or enabled via a focused effort, and then conceptualize what that focused effort might look like.

Workshop participants and observers included representatives from academia, government agencies, industry, and national laboratories and who conduct basic research (e.g., NIH- and DOE-funded) in development and application of technologies, methods, tools, and resources to enable small molecule measurements; conduct research to identify novel bioactive small molecules; perform molecular measurements to understand their roles in biological and environmental systems (also known as chemical ecology) in support of basic science research and national missions; lead government agency programs [e.g., Department of Homeland Security (DHS) and Department of Defense (DoD)] whose missions depend on small molecule measurements; develop novel materials or non-conventional technologies for small molecule measurements; develop or have expertise in computational chemistry, cheminformatics, high performance and cloud computing; or were instrumental in the stand up and execution of the HGP (**Figure 2; Appendix I**).

All are united by their implementation of similar analytical strategies in the form of 'metabolomics,' 'metabolic profiling,' 'lipidomics,' 'natural product discovery,' 'non-targeted analysis,' 'suspect screening,' 'exposomics,' and 'biological and chemical threat detection.'

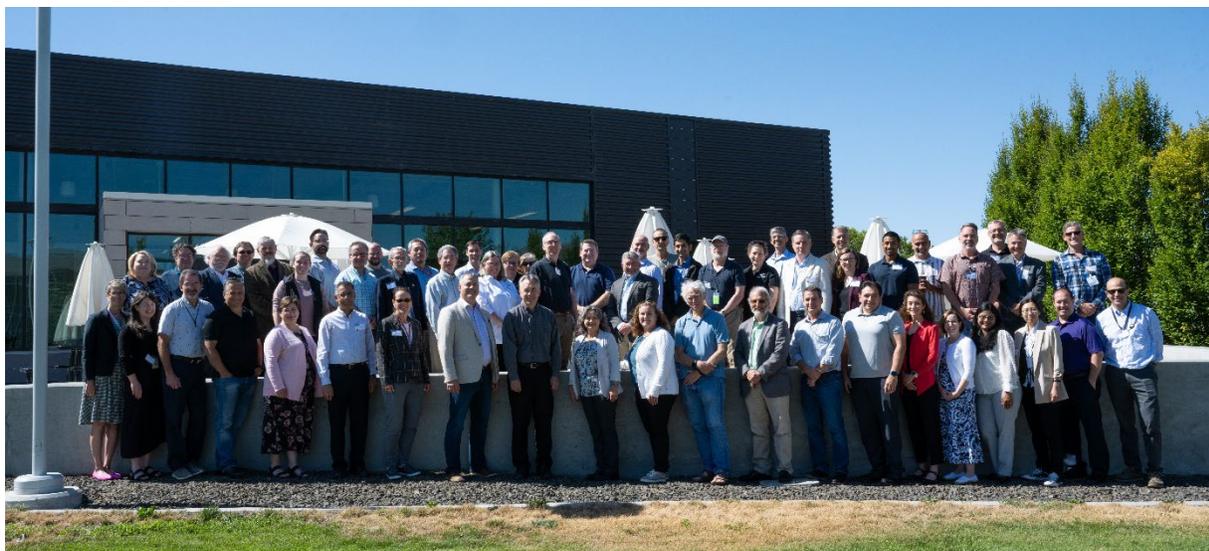

**Figure 2. Participants in the Decoding the Molecular Universe Workshop.** Sixty-two representatives from academia, government agencies, industry and national laboratories participated or were observers at the workshop. Not pictured are the following virtual attendees: Dr. Patrick Chain (Los Alamos National Laboratory); Dr. Kirsten Hofmockel (PNNL); Dr. Ann Knolhoff (U.S. Food & Drug Administration); Dr. Trent Northern (Lawrence Berkeley National Laboratory); Dr. Pablo Rabinowicz (U.S. Department of Energy (DOE) Office of Biological and Environmental Research (BER)); Dr. Paul Sammak (U.S. DOE BER); and Dr. Richard Smith (PNNL).

## WORKSHOP PURPOSE AND GOALS

Thomas Metz (PNNL) opened the workshop by welcoming participants and observers to the laboratory and providing a brief overview of the DOE national laboratory network and a historical background on PNNL. He then briefly reviewed the names and affiliations of workshop participants and observers and reminded the group of the diversity of basic research and applied mission areas represented.

Metz then reviewed the workshop purpose and goals, highlighting the 1980s workshops, led by Sinsheimer and Bitensky, on the feasibility of a human genome project as inspiration. He pointed out the seemingly disparate nature of the research and mission space of participants and observers, yet highlighted the similar approaches implemented by each (e.g., 'metabolomics,' 'suspect screening,' 'chemical/biological threat assessment,' etc.) towards the common goal of high-throughput, unambiguous identification and quantification of all small molecules in any sample relevant to the hypotheses or questions being addressed. It was noted that similarity in approach leads to similar technical challenges in achieving research and mission goals.

As such, one goal of the workshop was to bring the representatives from these areas together to discuss and ideate towards new solutions and ideally a step-function change in the field of small molecule measurement. Metz pointed out that the general analytical approach for performing high-throughput measurements of small molecules and correlating the results to, for example, a biological phenotype was essentially conceptualized by Linus Pauling and Arthur Robinson over 50 years ago as 'orthomolecular medicine'[4, 5] but had not significantly changed since.

Within that span of time, the scientific community galvanized around the concept of sequencing the human genome and accomplished that historical task. Finally, Metz noted that the single, most focused U.S. effort to date on metabolomics and small molecule identification was the NIH Common Fund Metabolomics Program, which invested ~$200 million over a 10-year program. That represented ~6.7% of the investment made towards sequencing the human genome, yet the complexity of small molecule chemical structures is much more complex than that of DNA and RNA. The concluding charge delivered to the group was to discuss the impact, feasibility, and cost of an engineering project like the HGP but focused on the molecular universe.

**HUMAN GENOME ROW**

An opening plenary lecture was jointly delivered by Drs. Ari Patrinos (Novim Group) and Dan Drell (U.S. DOE BER, retired) who provided a historical overview and perspective of the execution and the scientific and technical outcomes of the HGP, as well as key lessons learned.

> There are two kinds of scientific revolutions, those driven by new tools and those driven by new concepts…The effect of a concept-driven revolution is to explain old things in new ways. The effect of a tool-driven revolution is to discover new things that have to be explained. In almost every branch of science, and especially in biology and astronomy, there has been a preponderance of tool-driven revolutions. We have been more successful in discovering new things than in explaining old ones.
>
> Freeman Dyson, *Imagined Worlds*

Patrinos opened his half of the lecture – titled Human Genome Row in analogy to the Bob Dylan song Desolation Row – by quoting Freeman Dyson, who had highlighted two types of scientific revolutions – concept-driven scientific revolutions and tool-driven scientific revolutions – and the fact that the scientific community has had much more success with the latter than the former.

He pointed out that the field of biology historically had been concept-driven, but that what changed it into a tool-driven scientific revolution was the HGP. Dr. Patrinos highlighted the origins of the HGP, and that while many individuals were involved and take credit, it was Charles DeLisi who was the force behind the effort to launch the program. He also noted that the eventual collaboration between the DOE and NIH helped to overcome differing opinions among scientists and leadership in different agencies. The competition between the public-funded effort and a private sector effort led by J. Craig Venter (via Celera Genomics) ultimately helped drive the respective efforts towards completion of the sequencing of the human genome and eventually the two efforts became cooperative.

> What this underscores is, in many ways, in my view, one of the huge powers of the HGP…had we not initiated this project: the use of genome sequencing not as an end in and of itself but as a monitor of biological processes that are ongoing.
>
> Dan Drell

Dan Drell then spoke about the aspects of the HGP that allowed it to succeed. First and foremost, it was not a fundamental scientific research project but an infrastructure-building effort, with foundational prior knowledge; simple, universally agreed-on, and impactful metrics of success; explainable value to society; strong leadership and coordination; and strong, consistent, bipartisan political support.

Drell then provided his perspective on what the world would be like had the HGP never occurred, including: "orders of magnitude slower generation of sequence"; vastly less survey sequencing to capture the dynamics of human (and other) genomes; significantly reduced appreciation for the proportion of the vertebrate genome that was dedicated to non-coding genes; much less impetus for development of computational biology; and many fewer genetic tests.

His guidance for a small molecule-focused effort was not to do things in isolation, but to leverage the talents and infrastructures that exist broadly in the scientific community; not to wait until the end to address any potential ethical issues that might arise; to be sure that the work aligns with the supporting funding agencies; and not to over promise yet to think big.

**WORKSHOP PROCESS: KEY QUESTIONS, DISCUSSION, AND RESPONSE THEMES**

Three Key Questions were considered, with three lightning talks presented in response to each question with diverse perspective, in order to stimulate discussions that subsequently occurred during breakout sessions.

Participants and observers were then assigned to 4 breakout groups, where the Key Questions were considered and discussed. Breakout group discussion leads then summarized the main points from their discussion groups and reported these back to the main group.

Key Themes were identified from among breakout group discussions of each key question and were summarized and documented. Key Themes were defined as those that were reported by ≥50% of breakout groups, although many were reported by 75-100% of groups.

| Key Question | Key Theme/s Identified |
|---|---|
| 1. What would be possible if we could determine the complete chemical composition of any sample? | 1.1.   Complete knowledge and/or complete understanding of a system.<br><br>1.2.   Ability to design and/or manipulate systems for desired outcomes |
| 2. How far can existing technologies get us? | 2. 1. The analytical and computational tools and chemical knowledge are available now to implement fully standardized methods and open-data exchange formats in order to identify a manageable portion of the small molecule universe, but the adoption of data standards and the integration of data into repositories with full sample metadata are required. |

|  | 2.2. AI/ML-based computational methods for faster small molecule identification timescales are available now, but these need to be automated and made more robust by leveraging extensive, diverse, and high confidence training data. |
|  | 2.3. All available small molecule reference libraries should be increased, pooled, and made publicly available. |
|  | 2.4. Analytical techniques for acquiring small molecule data should be combined into multi- dimensional or multi-modal methodologies. |
| 3. Which instrumental, computational, or other advancements need to occur? | 3. 1. Systematic aggregation, curation, validation, and translation of existing small molecule reference databases. |
|  | 3. 2. Accurate and confident computational predictions of small molecule properties for in silico annotation using validated tools that report accuracies and ranges of uncertainty. Single standalone instrument with integrated orthogonal measurements that will operate without reliance on small molecule reference libraries. |

**Key Question 1:** *What would be possible if we could determine the complete chemical composition of any sample?*

Lightning talks were presented in response to Key Question 1 by researchers from diverse disciplines: basic science of human exposures and exposomics (Gary Miller, Columbia University), emerging chemical and biological threats to the modern warfighter (Morgan Minyard, U.S. Army DEVCOM CBC), and environmental microbial metabolism and microbial communities (Patrick Chain, Los Alamos National Laboratory).

> What we have now could be influential, but every step toward this meeting's premise will represent saltatory leaps in biological understanding.
>
> Gary Miller

Gary Miller placed boundaries on the response to the question by asking "What would be possible if we could unequivocally identify all the peaks from liquid chromatography, or gas chromatography high-resolution mass spectrometry from human plasma samples?" Miller described the typical approach for performing small molecule analysis in the context of 'exposomics' studies and answered that all features detected in high-throughput assays would no longer be mysterious but would become identified small molecules, from endogenous and anthropogenic origins. We would also know the proximate molecular causes of any disease (and therefore be able to leverage genetics). He emphasized that complete knowledge of the small molecule composition of any sample may reveal truths that the general public and government are not yet ready to handle (e.g., that living in a particular city may increase vulnerability to disease).

Morgan Minyard introduced the mission space of threat agent science and the types of questions and decisions used to prioritize small molecules of interest (i.e., chemical threats) for further study and characterization, in order to frame her response to the question. She highlighted organ-on-a-chip technologies (liver, lung, brain, heart, kidney) for predictive toxicology as one of the core technologies to gain knowledge about breakdown products and metabolites. A key part of that process was a reliance on in silico approaches for predictive toxicology since there is a long list of chemical threats that are currently being encountered and that might be encountered in the future. In response to the question, Minyard pointed out that it would be possible to identify breakdown products of threat agents that may also be toxic, in addition to the agents themselves.

Patrick Chain provided a perspective from a microbiome/genomics scientist and highlighted the role of microbiomes in further metabolizing, detoxifying, etc. xenobiotic and anthropogenic small molecules. In response to Key Question #1, Chain answered that we would learn everything and anything about a system, including the function of ~30-60% of genes that remain uncharacterized in function for a given microbiome and the identity of 50-80% of all metabolites that are currently unknown. He reiterated the complexity of small molecule space and the diversity in methods used to measure it, and that multi-omics approaches (e.g., 'proteomics', 'metabolomics', 'lipidomics') to understanding systems will be key.

**Key Theme 1.1:**

***Complete knowledge and/or complete understanding of a system.*** This includes biological systems in general, but for human systems in particular the small molecule basis of all disease and complete knowledge of the totality of lifetime exposures to chemicals. During group discussion of this key theme, Oliver Fiehn (University of California, Davis) stressed the importance of having clearly defined endpoint phenotypes in order to appropriately place bounds on the interpretation of the knowledge of a system, as well as the importance of not just focusing on the parts lists of systems but also elucidation of the interactions between the parts of the system.

David Balshaw (National Institute of Environmental Health Sciences) emphasized that a key result of attaining complete knowledge or understanding of a system would be causality. Roger Linington (Simon Fraser University) stated that achieving the outcomes of this key theme would fundamentally change sub-disciplines and fields, for example in the field of natural products discovery where >80% of effort is on wet lab work. There was some discussion about trying to define which sample(s) and which analytical metrics should be considered when trying to answer Key Question 1; however, others preferred to discuss these points later in the workshop. Trent Northen (Lawrence Berkeley National Laboratory) described how our current lack of knowledge of plant, microbial, and soil small molecules not only limits our understanding of genomics but also soil nutrient and carbon cycling, both of which are central to sustainable agriculture and bioenergy.

A comprehensive understanding of environmental small molecules would, for example, enable harnessing beneficial microbes to reduce fertilizer use and help restore depleted soil carbon stocks. It would also greatly enable a new predictive understanding of environmental microbiomes to identify tipping points in community stability and how changing community compositions will impact key biogeochemical cycles. Building on this, Chris Henry (Argonne National Laboratory) indicated how comprehensive data on observed and identified metabolites would revolutionize functional genomics when combined with methods that identify enzymes and species present in a given sample. We would be able to systematically determine how observed small molecules interact with organisms and enzymes, elucidate the small molecule transformations mediated by enzymes, and thus massively advance efforts to decode the mechanistic functions of enzymes across the tree of life.

**Key Theme 1. 2:**

***Ability to design and/or manipulate systems for desired outcomes.*** This includes the design of sustainable and fit-for-purpose systems, such as for bioeconomic purposes, the transformation of agricultural practices via development of new fertilizers, manipulation of plant-microbe interactions, etc. to improve sustainability, and the ability to increase carbon sequestration in soil in order to mitigate the effects of climate change. This could be further applied beyond Earth, such as by facilitating exploration of life on other planets, mining chemicals on other worlds for sustainable space exploration, and monitoring crew health during missions. There was no additional group discussion on this key theme.

**Key Question 2:** *How far can existing technologies get us?*

Lightning talks were presented in response to Key Question 2 by researchers conducting basic and applied research in development and application of non-targeted analysis in food safety applications (Christine Fisher (O'Donnell), U.S. Food and Drug Administration (FDA)), discovery and characterization of novel bioactive small molecules (Neha Garg, Georgia Institute of Technology), and development and application of novel computational tools and resources for measuring and studying small molecules (David Wishart, University of Alberta).

Christine Fisher described the challenges faced by the FDA when analyzing foods for safety purposes, including a variety of food types, a variety of food brands, and the chemical complexity and number of small molecules that can be detected during analyses. She then stated three primary goals for food safety analysis: detect as many small molecules as possible, prioritize/reduce the data, and understand data quality and the limitations of the associated methods and tools.

Fisher argued that the availability and analysis of pure reference compounds are critical for achieving these goals and presented her group's efforts to develop and work toward a broadly available standard mixture that could be used by the non-targeted community for a variety of quality control purposes. In response to Key Question 2, Fisher stated that existing technologies do not adequately provide unambiguous identifications of all small molecules on a routine timescale but qualified her response by reiterating that this is often not required for food safety purposes.

Neha Garg then provided an overview of machine learning- based methods developed by the community for providing insights into the chemistries of the

'dark metabolome,' including molecular networking, MS2LDA,[21] CANOPUS,[22] MassQL,[23] and MASST[24] and showed representative examples of successful implementations of each tool in the identification and chemical classification of novel natural products and annotation of microbially-derived novel antibiotic biotransformation products.

Garg highlighted that these methods have all been developed in the last 10-12 years, some of which are still in the bioRxiv and not yet in peer-reviewed, published forms.

> Even with the advanced methods that use machine learning and in silico prediction, we can still only annotate 10% of the metabolome.
>
> Neha Garg

She presented a case study reflecting on how application of these methods advanced the mechanistic understanding of personalized interactions between trimethoprim and trimethoprim-resistant microbial pathogens, resulting in an improved understanding of treatment failure. In response to the question under consideration, Garg emphasized that tools for machine learning-based in silico predictions of measured molecular properties, such as annotations based on predicted tandem mass spectra and chemical classification, although promising, are in their infancy.

The last speaker to present a lightning talk in response to Key Question 2 was David Wishart. Wishart first provided definitions of the terms 'chemicals' and 'detectable' to frame the context of his presentation. He defined 'chemicals' as small molecules less than 1500 Da, including along the spectrum of organic to inorganic, as well as synthetic and naturally occurring compounds but excluding synthetic and natural polymers such as proteins, polysaccharides, DNA/RNA, cross-linked polymers, etc.

> My best estimate, based on the number of 2 million known compounds, is that there are at least 25 million unknown compounds that are at levels that would have some biological or environmental influence.
>
> David Wishart

By 'detectable', he referred to compounds that were present at levels that could have a biological effect, defined as 1 part per trillion (e.g., the level of detection of grapefruit mercaptan). Wishart then reviewed the technologies typically employed in high-throughput analysis of small molecules and their ranges of sensitivities. He then reviewed the numbers of known chemicals reported in decreasing order in online resources such as PubChem,[25] the EPA's DSSTox database,[26] COCONUT,[27] PubChem Lite,[28] HMDB,[29] and the NORMAN SLE,[30] that totaled around 2.1 million non-redundant chemicals that are known or are knowable.

Finally, Wishart suggested that the number of small molecules that can be routinely measured is a few thousand using current technologies and approaches. Wishart noted that the instrumentation and technology does exist to detect small molecules at parts per trillions levels, and with heroic efforts it may be possible to collect information on 1 million or more compounds in a given sample. However, the community does not have the authentic standards or reference information to positively identify those small molecules, and he posited that the community will never have such numbers of authentic standards. As an alternative, he suggested the community transition to *in silico* or computationally predicted chromatographic, spectral or structural reference information for enabling unknown small molecule identification. He highlighted the success of community efforts to date while pointing out that the accuracies of such predictions need to improve by 10-20% to achieve the desired level of success in novel compound identification.

**Key Theme 2. 1:**

The analytical and computational tools and chemical knowledge are available now to implement fully standardized methods and open-data exchange formats in order to identify a manageable portion of the small molecule universe, but the adoption of data standards and the integration of data into repositories with full sample metadata are required.

**Key Theme 2. 2:**

AI/ML-based computational methods for faster small molecule identification timescales are available now, but these need to be automated and made more robust by leveraging extensive, diverse, and high confidence training data.

**Key Theme 2. 3:**

All available small molecule reference libraries should be increased, pooled, and made publicly available.

**Key Theme 2. 4:**

Analytical techniques for acquiring small molecule data should be combined into multi-dimensional or multi-modal methodologies. There was no additional group discussion.

**Key Question 3:** *Which instrumental, computational, or other advancements need to occur?*

Lightning talks were presented by researchers with applied expertise in chemical, biological, radiological, nuclear and explosives sensing (Augustus 'Way' Fountain, University of South Carolina), who perform development and application of new tools for small molecule assays (Facundo Fernandez, Georgia Institute of Technology), and who apply molecular measurements in exploring emerging environmental contaminants and their health implications (Kirsten Overdahl, National Institute of Environmental Health Sciences).

> There are literally thousands of lethal materials that can be used, and we should not just be concerned with those prohibited by the Chemical or Biological Weapons Treaties.
>
> Way Fountain

Augustus 'Way' Fountain introduced the area of remote sensing of hazardous materials and the challenges of assigning attribution to strategic events in operational fields. He pointed out that field detection is a critical need, because of lengthy turnaround times associated with sending a field sample back to the U.S for laboratory-based analyses and reporting. He highlighted a 1985 commentary in the journal Science by Thomas Hirschfield,[31] who predicted increasing hyphenation of instrumentation, the utility of orthogonal techniques, and a trend toward miniaturization in future small molecule analysis.

Fountain indicated that moving towards improved mobile measurements will be needed for future operational environments, and that focus should be particularly made to the various aspects of sampling. Finally, Fountain suggested that distributed sampling, even if it consists of relatively low resolution measurements, will be valuable in field detection.

Facundo Fernandez described the NIH Common Fund program Molecular Transducers of Physical Activity Consortium (MoTrPAC) and illustrated the small percentage of small molecules that can be confidently identified using conventional molecular analysis techniques. He then reviewed the concept of 'analytical peak capacity', which is defined as the number of real components of a sample that can be separated without overlap, and described the benefits achieved from coupling orthogonal techniques in hyphenated measurement approaches.

One such hyphenated measurement platform is high performance liquid chromatography coupled with ion mobility spectrometry and high-resolution mass spectrometry (LC-IMS-MS). Fernandez then highlighted the benefits of ion mobility separations and the physicochemical property of collision cross section as an additional property for annotating small molecules and the use of shift reagents (such as crown ethers). He then described a typical workflow for performing small molecule measurements and annotating compounds using LC-IMS- MS. As a response to Key Question 3, he described a new instrument, the 'Frankentrometer,' which is an idealized and to-be-developed instrument for molecular measurements. His belief was that the majority of the future technical developments that would enable the Frankentrometer are needed at the front of the instrument, such as new chromatographic and molecular ionization advancements. However, he pointed out that there is a relative reduction of instrument development funding, and concomitantly, effort at universities.

Kirsten Overdahl provided the final lightning presentation in response to Key Question 3. She opened by reframing the question as 'What instrumental or computational abilities currently hinder our abilities to perform effective horizon scanning and achieve a comprehensive assessment of the toxome?" Overdahl then reviewed a typical analytical pipeline for determining sample small molecule composition and consequence, and identified the step of chemical annotation (including re-identification of known small molecule space) as a current bottleneck.

She suggested computational strategies that resolve small molecule annotation through analogy-based approaches as a promising area that should be further explored. Specifically, Overdahl asserted that even if de novo annotation is not yet universally possible, analogy-based approaches to date (such as those enabled by tools and resources such as BinDiscover[32] and molecular networking[33]) have successfully shown the benefits of using comprehensive spectral descriptors to understand how small molecules fit into relationships with each other, helping each to provide insight on molecule-specific information and to better understand the scope of small molecule space. Overdahl suggested directions in which to expand these tools and proposed that collective thought

> We have precedent for the first time ever of the National Academies of Science recommending in official documentation the use of sub-class approaches for regulation, and I believe we should interpret this as our green light to pursue both analogy-based annotation to define those subclasses and better structure-based annotation to gain better physical-chemical property insight into those subclasses.
>
> Kirsten Overdahl

is needed about how to harmonize metadata reporting in a way that accounts for matrix-specific and endpoint-specific details and works for both endogenous and exogenous small molecules.

**Key Theme 3. 1:**

Systematic aggregation, curation, validation, and translation of existing small molecule reference databases. This theme was described as facilitating a better understanding of data provenance, the use of more sustainable databases, and the application of more orthogonal techniques. During group discussion, it was identified as potentially very low hanging fruit in terms of an innovation that could be rapidly implemented.

**Key Theme 3. 2:**

Accurate and confident computational predictions of small molecule properties for in silico annotation using validated tools that report accuracies and ranges of uncertainty. It was pointed out that this area may require a significant investment in national computational and data resources. In several Key Theme discussions across the Key Questions, data curation and management was a critical condition for success. This includes measured, simulated, and predicted data that is aggregated and managed across the community. This theme was also emphasized in a recent Nature editorial titled "For chemists, the AI revolution has yet to happen".[34]

**Key Theme 3. 3:**

Single standalone instrument with integrated orthogonal measurements that will operate without reliance on small molecule reference libraries. This objective is particularly necessary for field deployable applications, and includes non-conventional measurements made in high-throughput paradigms, such as spectroscopic (e.g., infrared, microwave, etc.) methods. It was noted that there currently does not seem to be as much emphasis placed on instrumentation development as in previous decades.

**WORKSHOP WRAP-UP - ON CONVERTING AN IDEA INTO POLICY: THE ORIGIN OF THE HUMAN GENOME PROJECT**

Charles DeLisi (now at Boston University) provided a final (recorded) plenary lecture that described the initiating events and intra- and interagency communications (primarily led by his significant efforts), together with significant political support, that resulted in the HGP. He began his presentation by highlighting that the concept for the HGP originated in an odd place – the U.S. Department of Energy – however, he pointed out that at the time the DOE had the necessary culture in which a large complex project could take root and named David Smith, Alvin Trivelpiece, and William Flynn Martin as key supporters of moving the HGP concept to policy within that organization.

> As a physics organization, DOE had a technological and managerial culture in which a large complex project could naturally take root.
>
> Charles DeLisi

Politically, he highlighted support and collaboration from then New Mexico Senator Pete Domenici, as well as from Judy Bostock and Tom Palmieri in the White House Office of Management and Budget, the latter of whom readily recognized that the HGP was not an open-ended science project but an engineering project that would build critical infrastructure that would support science – it had a beginning, a middle, and an end – and that was the key to eventually receiving broad support for the concept. However, he did highlight various aspects of political resistance, as well as resistance from the scientific community. DeLisi concluded his lecture with guidance for the participants if they are to pursue a similar effort and provided a general outlook of what scientific advances might be enabled if such a project were to proceed. He identified these as a "trifecta of importance": climate change, agriculture, and new medicines. These science challenges would be impacted not directly but indirectly due to a deeper understanding of the cell and human physiology, and translation would be quicker than achievable without it.

## A NATIONAL AND DIRECTED EFFORT

The workshop's concluding discussion focused on establishing one or more conceptual designs for a national, multi-institution effort to decode the molecular universe. Thomas Metz (PNNL) started by framing the discussion in the context of comparatively slow evolution of technologies for small molecule identification and the potential future where a minimally changed state-of-the-art would be a barrier to critical scientific discoveries. He challenged attendees to consider two technical approaches for decoding the molecular universe: 1) an approach based on scaling alone, for example massive parallelization of conventional methods that utilize fractionation, isolation, and identification of small molecules using traditional structure elucidation approaches, such as NMR spectroscopy, X-ray crystallography, or micro-electron diffraction or 2) higher throughput (e.g., mass spectrometry-based) approaches coupling significant innovations in data processing/analysis with the co-development of new "hyper-orthogonal" instrumentation and computational predictions of small molecule properties. Josh Adkins (PNNL) suggested that the two options were not mutually exclusive, but perhaps the project could begin right away with conventional approaches and also include parallel efforts in disruptive technologies or approaches. That suggestion garnered support from others.

In addition to modeling the project after the HGP, David Wishart (University of Alberta) strongly urged that the Protein Structure Initiative (PSI) be considered as an example. He pointed out that the single greatest barrier to decoding the molecular universe is a method for rapidly and confidently identifying small molecules in a high-throughput manner. A similar technical challenge was faced prior to launching the PSI, but that project resulted in the generation of significant infrastructure and generated much foundational data (including a well-defined protein fold universe) that, in turn, enabled highly innovative computational approaches for accurately predicting protein structure in high throughput (e.g., AlphaFold and RosettaFold). Wishart suggested that a similar effort focused on small molecules could provide tens of thousands of pure compounds for all to study using current conventional approaches, as well as compound characterization technologies both currently in development and yet to be developed.

Discussion then moved from the technologies and approaches to be implemented in such a project to what the target organism(s) or sample type should be. Roger Linington (Simon Fraser University) suggested to start not with a target organism but with a defined small molecule space that is organism independent. He stated that there is likely a finite set of molecular scaffolds in the small molecule universe that could be used as an initial target set and that would be representative of the majority of chemistries found in nature. The chosen small molecule set would provide researchers using a variety of technologies and approaches an opportunity to test the bounds of their approaches. It would also be a physical resource for researchers to use for benchmarking purposes in the years to come.

Linington's recommendation was met with broad support from the group. Katherine Duncan (University of Strathclyde) suggested that some or all of these molecular scaffolds might be 'keystone small molecules' that might serve key functions in various biological and environmental systems.

Before any project could begin, Oliver Fiehn (University of California, Davis) stressed the importance of establishing a systematic and well-engineered framework for collecting and processing of data so that future technologies would be able to re-use existing data. This would make it possible to seamlessly integrate newly generated data. For example, kits of internal standards and NIST pooled quality control samples (e.g., NIST SRM 1950) are already available to link systematically generated data from today to data contained in new repositories tomorrow.

Discussion then moved to the ultimate target of the project in terms of sample type(s) or organism(s) that would be studied after the defined set of molecular scaffolds. It was broadly agreed that a 'hook' was needed to excite the scientific community and the public. Some thought that the hook could be as simple as completing the knowledge of the central dogma of biology, since it is currently incomplete without a fuller understanding of small molecules and their roles in biological and environmental systems.

Others suggested that a focus on the 'exposome' and biological and environmental exposures would be an exciting target. Jon Sobus (U.S. Environmental Protection Agency) suggested that there could simply be one overarching hook or target (e.g., decoding the molecular universe) but several sub-targets that would all then benefit from the primary target. One example suggested was the National Emerging Contaminants Research Initiative sponsored by the White House Office of Science and Technology Policy, which in part is focused on characterizing emerging contaminants in drinking water. A last target that garnered discussion was a concept put forth by Chris Henry (Argonne National Laboratory) of the 'Chemical Tapestry of America', where the small molecule composition of every square mile of the country would be determined, although there were concerns that may be too large an effort. Recent projects already underway using wearable devices offer a tantalizing approach for tackling the sample collection aspects of this challenge.

Support ultimately formed around a project titled 'The Chemistry of Life Project' and that the breadth of such a project would include many other efforts that may go by other names or involve other concepts that are more esoteric. There was also the view that the various aspects of 'The Chemistry of Life Project' would be attractive to the broader public audience via tailored explanations. There was some concern voiced that the word 'chemistry' may evoke the making and breaking of bonds, rather than chemical structures of small molecules, so perhaps the word 'molecules' or 'metabolism' could be considered in place of 'chemistry'. In terms of a target sample or target organism, there was mixed opinion on whether to keep the target more general (e.g., on chemistry or small molecules) or to target a specific organism such as humans in order to have the broadest public support. There was eventual support for studying organisms of increasing complexity, culminating in humans, just as it occurred during the HGP.

The group then defined 'The Chemistry of Life Project'.

**The Chemistry of Life Project**

**Thrust 1:** Establish a systematic and well-engineered framework for collecting and processing data from the analysis of small molecules so that both current and future essential next generation technologies would be able to reuse existing data and also so that it would be possible to seamlessly integrate newly generated data.

**Thrust 2:** Assemble a well-informed and curated library of molecular scaffolds that could be used as an initial target set and that would be representative of all small molecule chemistries found in nature.

The chosen molecular set would provide researchers using a variety of current and future next generation technologies and approaches an opportunity to test the bounds of their approaches and would also be a physical resource for researchers to use for benchmarking purposes in the years to come.

Critically, this library would also enable the training of models to enable the development of predictive computational tools for disciplines such as metabolomics to solve the small molecule structure problem.

**Thrust 3:** Develop next generation experimental and computational technologies for small molecule analysis and apply these as developed and concurrently with current technologies and approaches in the complete molecular characterization of minimally complex organisms or systems, such as a bacterium or insect.

**Thrust 4:** Determine the complete small molecule composition of a composite human reference sample.

**FUTURE ACTIONS**

We as a founding coalition of experts will continue to discuss and refine the Chemistry of Life Project concept, and we encourage government agencies and the scientific community to also consider the value and impact of such a project towards filling remaining knowledge and technology gaps in human disease, biological and environmental systems, and threat molecular assessment. Such consideration can take the form of additional ad hoc workshops or formally commissioned activities, such as study by the National Academies.

# APPENDIX I

# WORKSHOP AGENDA

# AGENDA

## Decoding the Molecular Universe Workshop

Discovery Hall, Pacific Northwest National Laboratory, Richland, WA
Wednesday, August 9th through Thursday, August 10, 2023 | 7:30 am – 5:00 pm

### Day 1 | August 9

| TIME | TOPIC | LOCATION | PARTICIPANT/LEAD |
| --- | --- | --- | --- |
| 7:30 – 8:30 am | Badging, Check-In/Registration & Networking | Discovery Hall/ Badging Office | Karli Taxdahl<br>Jodi Giles |
| 8:30 – 8:45 am | Welcome & Logistics | Discovery Hall/ Room C | Tom Metz |
| 8:45 – 9:00 am | Workshop Purpose & Goals | Room C | Tom Metz |
| 9:00 – 10:00 am | Plenary – The Human Genome Project: Enabling S&T, Logistical Execution, and Impacts to Biological Understanding | Room C | Ari Patrinos<br>Dan Drell |
| 10:00 – 10:10 am | Break | | |
| 10:10 – 11:00 am | Key Question #1 – What would be possible if we could determine the complete chemical composition of any sample?<br>• Lightning Talk 1: 10:10 – 10:25<br>• Lightning Talk 2: 10:25 – 10:40<br>• Lightning Talk 3: 10:40 – 10:55 | Room C | Bobbie-Jo Webb-Robertson<br><br>Gary Miller<br>Morgan Minyard<br>Patrick Chain |
| 11:00 – 12:00 pm | Key Question #1 Breakouts<br>• Group 1<br>• Group 2<br>• Group 3<br>• Group 4 | <br>Room C<br>Room D<br>Room E<br>Room Frontier | David Balshaw<br>Katrina Waters<br>Kabrena Rodda<br>Tobias Kind |
| 12:00 – 1:30 pm | Key Question #1 – Report Out & Discuss (Working Lunch) | | David Balshaw<br>Katrina Waters<br>Kabrena Rodda<br>Tobias Kind |
| 1:30 – 2:20 pm | Key Question #2 – How far can existing technologies get us?<br>• Lightning Talk 1: 1:30 – 1:45<br>• Lightning Talk 2: 1:45 – 2:00<br>• Lightning Talk 3: 2:00 – 2:15 | Room C | Yehia Ibrahim<br><br>Christine Fisher<br>Neha Garg<br>David Wishart |
| 2:20 – 2:30 pm | Break | | |

| Time | Topic | Location | Participant/Lead |
|---|---|---|---|
| 2:30 – 3:35 pm | Key Question #2 Breakouts | | |
| | • Group 1 | Room C | Oliver Fiehn |
| | • Group 2 | Room D | Roger Linington |
| | • Group 3 | Room E | Rachel Gooding |
| | • Group 4 | Room Frontier | Kate Duncan |
| 3:35 – 4:50 pm | Key Question #2 – Report Out & Discuss | | Oliver Fiehn<br>Roger Linington<br>Rachel Gooding<br>Kate Duncan |
| 4:50 – 5:00 pm | Summary of Day 1 | Room C | Tom Metz |
| 6:00 – 8:00 pm | Dinner & Social | Tagaris<br>844 Tulip Lane<br>Richland, WA 99352 | |

## Day 2 | August 10

| TIME | TOPIC | LOCATION | PARTICIPANT/LEAD |
|---|---|---|---|
| 8:00 – 8:15 am | Networking | Discovery Hall | All |
| 8:15 – 8:25 am | Day 1 Recap | Room C | Tom Metz |
| 8:25 – 9:15 am | Key Question #3 – Which instrumental, computational, or other advancements need to occur?<br>• Lightning Talk 1: 8:25 – 8:40<br>• Lightning Talk 2: 8:40 – 8:55<br>• Lightning Talk 3: 8:55 – 9:10 | Room C | Simone Raugei<br><br>Way Fountain<br>Facundo Fernandez<br>Kirsten Overdahl |
| 9:15 – 10:20 am | Key Question #3 Breakouts | | |
| | • Group 1 | Room C | Jon Sobus |
| | • Group 2 | Room D | Marianne Sowa |
| | • Group 3 | Room E | Justin Teeguarden |
| | • Group 4 | Room Frontier | Deep Jaitly |
| 10:20 – 10:30 am | Break | | |
| 10:30 – 11:45 am | Key Question #3 – Report Out & Discuss | | Jon Sobus<br>Marianne Sowa<br>Justin Teeguarden<br>Deep Jaitly |
| 11:45 – 1:00 pm | Plenary – The Origin of the Human Genome Project: On Converting an Idea into Policy (Working Lunch) | Room C | Charles DeLisi (Recorded) |
| 1:00 – 2:15 pm | Is a National/International and Directed Effort Needed? | Room C | All |
| 2:15 – 2:25 pm | Break | | |
| 2:25 – 4:20 pm | Outline Position Paper | Room C | All |
| 4:20 – 4:50 pm | Summary of Workshop | Room C | Tom Metz |

# APPENDIX 2

# PARTICIPANT / OBSERVER LIST

Workshop Participants

| | |
|---|---|
| Josh Adkins | Biological Sciences Division, Earth and Biological Sciences Directorate, Pacific Northwest National Laboratory, Richland, WA, USA |
| Peter B. Armentrout | Department of Chemistry, University of Utah, Salt Lake City, UT, USA |
| David Balshaw | Division of Extramural Research and Training, National Institute for Environmental Health Sciences, Durham, NC, USA |
| Nelson Vineuza Benitez | Department of Textile Engineering, Chemistry, and Science, North Carolina State University, Raleigh, NC, USA |
| Evan Edward Bolton | National Center for Biotechnology Information, Bethesda, MD, USA |
| Patrick Chain | Bioscience Division, Los Alamos National Laboratory, Los Alamos, NM, USA |
| Fanny Chu | National Security Directorate, Pacific Northwest National Laboratory, Richland, WA, USA |
| Court Corley | National Security Directorate, Pacific Northwest National Laboratory, Richland, WA, USA |
| John Cort | Biological Sciences Division, Earth and Biological Sciences Directorate, Pacific Northwest National Laboratory, Richland, WA, USA |
| Donald Cronce | Advanced and Emerging Threat Division, US Defense Threat Reduction Agency, Fort Belvoir, VA, USA |
| Emily Davis | Department of Defense |
| Elizabeth Denis | National Security Directorate, Pacific Northwest National Laboratory, Richland, WA, USA |
| Daniel Drell | Office of Biological and Environmental Research, U.S. Department of Energy, Washington, DC |
| Katherine Duncan (Kate) | University of Strathclyde, Strathclyde Institute of Pharmacy and Biomedical Sciences, Glasgow, UK |
| James Evans | Environmental Molecular Science Division, Pacific Northwest National Laboratory, Richland, WA, USA |
| Robert Ewing | National Security Directorate, Pacific Northwest National Laboratory, Richland, WA, USA |
| Oliver Fiehn | University of California Davis, West Coast Metabolomics Center, Davis, CA, USA |
| Facundo Fernandez | School of Chemistry and Biochemistry and Petit Institute for Bioengineering and Bioscience, Georgia Institute of Technology, Atlanta, GA, USA |
| Christine Fisher (O'Donnell) | Center for Food Safety and Applied Nutrition (CFSAN), U.S. Food and Drug Administration (FDA), College Park, MD, USA |
| Augustus Way Fountain | University of South Carolina, Columbia, SC, USA |
| Neha Garg | School of Chemistry and Biochemistry and Petit Institute for Bioengineering and Bioscience, Center for Microbial Dynamics and Infection, Georgia Institute of Technology, Atlanta, GA, USA |
| Rachel Gooding | Hazard Awareness and Characterization Center, Technology Centers Division, Department of Homeland Security S&T, Chemical Security Analysis Center, Office of National Laboratories, Washington D.C., USA |
| Stefan Grimme | Mulliken Center for Theoretical Chemistry, University of Bonn, Bonn, Germany |
| Chaitanya Gupta | Probius Inc, Palo Alto, CA, USA |
| Chris Harrilal | Biological Sciences Division, Earth and Biological Sciences Directorate, Pacific Northwest National Laboratory, Richland, WA, USA |
| Jedidiah Hastings | Department of Defense |
| Chris Henry | Data Science and Learning Division, Argonne National Laboratory, Lemont, IL, USA |
| Robert (Bob) Hettich | Biosciences Division, Oak Ridge National Laboratory, Oak Ridge, TN, USA |
| Yehia Ibrahim | Biological Sciences Division, Earth and Biological Sciences Directorate, Pacific Northwest National Laboratory, Richland, WA, USA |
| Rabih Jabbour | CSAC Chemistry Security Laboratory, DHS, Edgewood, MD, USA |
| Navdeep Jaitley | Apple Machine Learning Research, Cupertino, CA, USA |
| Connor Jenkins | U.S. Army DEVCOM CBC Biodefense Division, Aberdeen Proving Ground, MD, USA |
| Tobias Kind | Enveda Biosciences, Boulder, CO, USA |
| Ann Knolhoff | Spectroscopy and Mass Spectrometry Branch, Food and Drug Administration's (FDA's) Center for Food Safety and Applied Nutrition, College Park, MD, USA |
| Roger Linington | Simon Fraser University, Burnaby, BC, Canada |
| Tom Metz | Biological Sciences Division, Earth and Biological Sciences Directorate, Pacific Northwest National Laboratory, Richland, WA, USA |
| Gary Miller | Mailman School of Public Health, Columbia University, New York, NY, USA |
| Morgan Minyard | U.S. Army Combat Capabilities Development Command Chemical and Biological Center (DEVCOM CBC), Aberdeen Proving Grounds, MD, USA |
| Trent Northern | Environmental Genomics and Systems Biology |
| | Division, UC Berkeley, Joint Genome Institute, Berkeley, CA, USA |
| Kirsten Overdahl | Metabolomics Core Facility, National Institute of Environmental Health Sciences, Research Triangle Park, NC, USA |
| Ari Patrinos | Novim Group, Santa Barbara, CA, USA |
| Dan Raftery | School of Medicine, University of Washington, Seattle, WAS, USA |
| Richard Smith | Biological Sciences Division, Earth and Biological Sciences Directorate, Pacific Northwest National Laboratory, Richland, WA, USA |
| Doug Sheeley | Office of Strategic Coordination, National Institutes of Health (NIH), Bethesda, MD, USA |
| Jon Sobus | Advanced Analytical Chemistry Methods Branch, Chemical Characterization and Exposure Division, Center for Computational Toxicology and Exposure, United States Environmental Protection Agency, Research Triangle Park, NC, USA |
| Justin Teeguarden | Environmental Molecular Sciences Laboratory, Pacific Northwest National Laboratory, Richland, WA, USA |
| Kabrena Rodda | National Security Directorate, Pacific Northwest National Laboratory, Richland, WA, USA |
| Paul Rigor | Center for Cloud Computing, Pacific Northwest National Laboratory, Richland, WAS, USA |
| Rachel Richardson | Biological Sciences Division, Earth and Biological Sciences Directorate, Pacific Northwest National Laboratory, Richland, WA, USA |
| Simone Raugei | Physical Sciences Division, Pacific Northwest, National Laboratory, Richland, WA, USA |
| Mariann Sowa | Science Directorate, NASA Ames Research Center, Silicon Valley, CA, USA |
| Akos Vertes | Department of Chemistry, George Washington University, Washington, DC, USA |
| Katrina Waters | Biological Sciences Division, Earth and Biological Sciences Directorate, Pacific Northwest National Laboratory, Richland, WA, USA |
| Bobbie-Jo Webb-Robertson | Biological Sciences Division, Earth and Biological Sciences Directorate, Pacific Northwest National Laboratory, Richland, WA, USA |
| Antony Williams | Computational Chemistry & Cheminformatics Branch, Chemical Characterization and Exposure Division, Center for Computational Toxicology and Exposure, United States Environmental Protection Agency, Research Triangle Park, NC, USA |
| David Wishart | Department of Biological Sciences, University of Alberta, Edmonton, AB, Canada |

**Workshop Observers**

| | |
|---|---|
| Pablo Rabinowicz | Biological and Environmental Research Program, Department of Energy (DOE), Washington, DC, USA |
| Paul Sammak | Biological and Environmental Research Program, Department of Energy (DOE), Washington, DC, USA |

# APPENDIX 3

# PARTICIPANT BIOS

# U.S. DOE NATIONAL LABORATORIES

**Joshua (Josh) Adkins** is a Laboratory Fellow, Co-Director of the Pacific northwest bioMedical Innovation Collaboratory (PMedIC – a joint institute with Oregon Health and Science University), and the PNNL Health and Human Services Sector Manager. He further leads large multi-disciplinary projects for the NIH and other federal sponsors. He is Director of the PNNL Proteomics Chemical Analysis Center for MoTrPAC (ProMoTr) and a LungMAP resource center. Dr. Adkins' research focuses on the comprehensive characterization of biomolecules through space (associated proteins, structural determinants, and localization) and time (before and after treatment, cell cycle, day-night cycle, and evolutionary changes) to better understand biological systems.

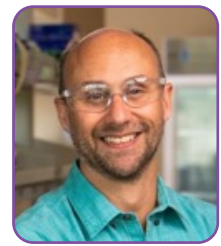

**Patrick Chain** is a Senior Scientist and Laboratory Research Fellow at Los Alamos National Laboratory's Bioscience Division. Dr. Chain manages a portfolio of diverse array of basic and applied research projects with a focus on microbes in both environmental and public health settings. He is also heavily involved in helping standardize various genomics approaches in these disparate scientific disciplines. He is co-PI on the National Microbiome Data Collaborative, a multi-national lab project whose goal is to help both standardize and link all publicly available microbiome omics data, including the various forms of both sequencing and mass spectrometry data. He is also keenly interested in linking various multiomics data and has been generating proteomic and metabolomic data to link them to genomic and transcriptomics data to better understand the dynamics between soil bacteria and fungi. While partial stories can be told with the analyzed data, a more comprehensive assessment of the metabolome would clearly provide a more complete story of these complex relationships.

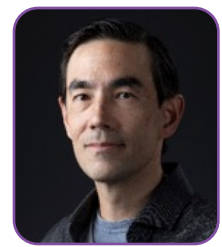

**Fanny Chu** is a data scientist in the Chem/Bio Informatics team in the National Security Directorate. Her research interests include development of analytical methods and data analysis tools to advance proteomics and metabolomics for chemical forensics and biodefense missions. She has recently worked on designing workflows to perform organism source characterization of unknown proteomics samples and on demonstrating the feasibility of human contaminant peptide marker detection in non-human samples for identification of sample preparers. Dr. Chu joined PNNL as a Post Doctorate Research Associate in 2020. She received her PhD in chemistry from Michigan State University in 2020, an M.S. in forensic science from Michigan State University in 2015, and a B.S. in chemistry from SUNY Binghamton in 2013. Her research has been published in Scientific Reports and Journal of Proteome Research. She has presented at the American Society for Mass Spectrometry, the American Academy of Forensic Sciences, and the Pittsburgh Conference on Analytical Chemistry and Applied Spectroscopy annual conferences.

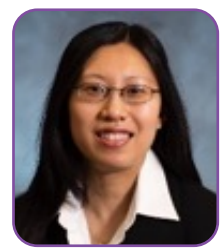

**Courtney Corley** is Chief Scientist for Artificial Intelligence at Pacific Northwest National Laboratory and has served in a variety of technical and line management roles during his tenure. His technical expertise is in the field of data science and biosurveillance for his foundational work in social media analytics, computational epidemiology, and natural language processing. His current research centers on developing machine learning approaches to solve challenges in limited data regimes, such as few-shot learning, and on trustworthy AI to assure AI systems are safe, secure, and robust. He is a mentor for the TechStars Industries of the Future startup accelerator, providing insight and guidance to founders leveraging data science and AI. Previously, Dr. Corley served as the Data Sciences and Analytics and Foundational Data Science group leader where his passion for talent and capability management has been an enabler and force-multiplier to meet the Nation's critical data science challenges. Dr. Corley co-led PNNL's Deep Learning for Scientific Discovery internal research investment that has applied deep learning across the breadth of PNNL's science and security missions. He holds a PhD in Computer Science and Engineering and an M.S. in Computer Science and is a member of IEEE, ACM, and AAAS.

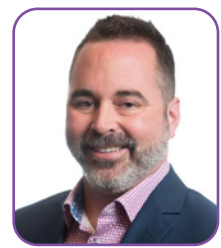

**John Cort** is a staff scientist in the Biological Sciences Division at Pacific Northwest National Laboratory (PNNL). Dr. Cort was an American Cancer Society postdoctoral fellow at PNNL from 1998-2000, and has been a PNNL staff member since 2001. Dr. Cort is also a joint faculty appointee in the Institute of Biological Chemistry at Washington State University. Dr. Cort is director of the Natural Products Magnetic Resonance Database (www.np-mrd.org), a database and repository for NMR data of natural products and specialized metabolites. Other current research interests include phenylpropanoid and lignin biosynthesis in plants, structure-property relationships in bio-oils derived from biomass, comparability determination in heterogeneous macromolecular pharmaceuticals and other complex mixtures, and protein and peptide structure-function-dynamics. Dr. Cort has also helped establish a role for NMR spectroscopy in chemical forensic source attribution and sample matching of toxic organic compounds and biomolecular toxins. Dr. Cort received a B.A. in Chemistry in 1991 from Williams College and a PhD in Organic Chemistry in 1997 from the University of Washington.

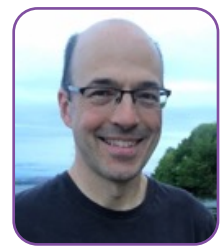

**Elizabeth Denis** is a Chemist at PNNL leading and supporting a variety of projects. Her work has focused on developing and optimizing methods for trace chemical detection of organic residues and vapors in the laboratory and from environmental samples using mass spectrometry. Over the last five years, Dr. Denis has further specialized in mass spectrometry and ion chemistry for trace detection and characterization of explosives and drugs. Many of her projects integrate laboratory experimental observations with computational modeling and statistics. Dr. Denis earned her PhD in Geosciences and Biogeochemistry from the Pennsylvania State University in 2016 and was an NSF Graduate Research Fellow. She earned a B.S. with honors in Geology-Chemistry from Brown University. Dr. Denis' passion for pushing the limits of our understanding of chemistry has driven her toward method development-focused research and has resulted in 2 patents and 20+ publications.

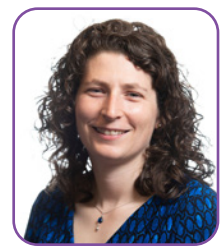

**James Evans** has been a Principal Investigator since 2009 with funding from NIH, DOE and internal PNNL/EMSL initiatives. His scientific interests focus on developing new methods and instrumentation to enable cryogenic, in-situ and dynamic multiscale and multimodal bioimaging. This includes a cell-free expression pipeline for in-vitro transcription, translation and 1-step purification of full-length proteins for downstream functional and structural studies using Native Mass Spectrometry and cryo-EM. Recently this pipeline solved the first structure of a protein called artemin in only 9 days after receiving the gene clone. This pipeline is now being adapted for true microscale structural and functional annotation of new proteins from only 25 microliter expression reactions that are split that for functional assays and direct structure determination with cryo-EM without purification. In addition to his science interests, he has been helping to expand access to cryo-EM more generally. Dr. Evans is PI of a DOE grant to provide cryo-EM/ET access to DOE researchers and is also M-PI on a NIH U24 funded project that is a joint venture with OHSU to build and operate the Pacific Northwest Center for Cryo-EM (PNCC) which has provided access to over 800 unique users in the last 4 years.

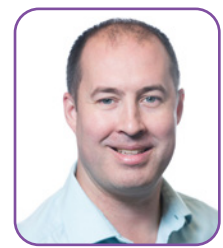

**Robert Ewing** is a staff scientist and team lead within the Chemical and Biological Signatures Group at Pacific Northwest National Laboratory (PNNL). Dr. Ewing is an analytical chemist receiving his PhD from New Mexico State University. Prior to joining PNNL in 2006, he was an Assistant Professor of Chemistry at New Mexico Institute of Mining and Technology (2002 to 2006), a Principal Scientist at the Idaho National Engineering and Environmental Laboratory (1997 to 2002), and a Senior Scientist with Geo-Centers Inc., Edgewood Area-Aberdeen Proving Grounds (1996 to 1997). He studies ion-molecule reactions at atmospheric pressure related to the detection of explosives, drugs, and other threat materials with mass spectrometry and ion mobility spectrometry. He developed the atmospheric flow tube-mass spectrometer that enables the vapor detection of drugs and explosives at parts-per-quadrillion levels. This technology was selected as a special recognition gold medalist for 2019 R&D 100 awards and a 2020 GeekWire award. Dr. Ewing has 6 US patents and over 45 peer reviewed publications. He has served on the board of directors for the International Society of Ion Mobility Spectrometry since 2006 and was President of that society from 2016 to 2018. He is currently the Chief Scientist of the PNNL *m/q* initiative.

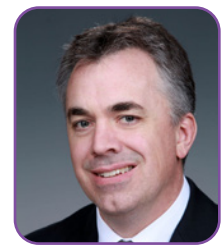

**Christopher Harrilal** is a chemist in the Environmental Molecular Sciences Laboratory here at PNNL. His current research is directed towards the further development of Structures for Lossless Ion Manipulations (SLIM) technology, a platform designed for ion transportation and mobility separations via traveling wave ion mobility. Additional research efforts have been directed towards the detailed modeling of ion-neutral collisions to explain the mobility difference observed during ultra-high resolution ion mobility separations. Dr. Harrilal received his PhD from Purdue University under the supervision of Professors Scott McLuckey and Tim Zwier. His graduate research involved using conformer specific ion spectroscopy for the structural elucidation peptide ions in the gas phase. The goal of this work was to understand the folding preferences of ions based on their primary sequence in a bottom-up approach. Dr. Harrilal has an expertise and significant background in the development of spectroscopic techniques, mass spectrometry, ion mobility, statistical mechanics, and molecular modeling.

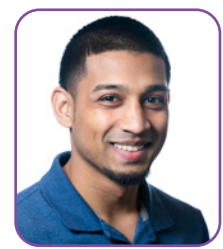

**Christopher Henry** is a group leader in the Data Science and Learning division of Argonne National Laboratory. He obtained his PhD in Chemical Engineering from Northwestern University, where he conducted research in biochemical thermodynamics, cheminformatics, and metabolic modeling. He went on to develop the ModelSEED pipeline for automated metabolic model reconstruction during his postdoc at Argonne. Today Dr. Henry is co-PI of the DOE Systems-biology Knowledgebase (www.kbase.us) project, leading development of a web-based collaborative platform for conducting large scale analysis and integration of genomic and multi-omics data. He continues to perform research exploring how metabolic models may be applied to integrate omics data to improve our understanding of gene function and microbial behavior. His team and collaborators are: (1) improving the quality of metabolic models with focus on sulfur, iron, methane, and nitrogen metabolism; (2) improving the quality of fungal metabolic models with focus on energy metabolism and lignin degradation; (3) predicting new metabolic pathways through the integrate of metabolomics data within metabolic models; (4) improving dynamic modeling of microbial communities with the goal of syncom design; and (5) predicting metabolic phenotypes for microbial genomes based on genomic data.

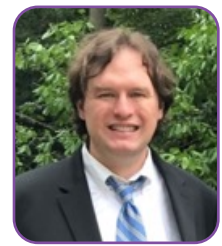

**Robert Hettich** is a distinguished research scientist and leader of the Bioanalytical Mass Spectrometry Group of the Biosciences Division at Oak Ridge National Laboratory and a joint faculty member in the Microbiology Department at the University of Tennessee. Dr. Hettich has over 35 years of experience in biological mass spectrometry, with a particular focus on high performance mass spectrometry. His research interests involve the development and application of advanced mass spectrometry technology for characterizing complex biological mixtures, such as microbial and plant proteomes. His group is active in the development of experimental and bioinformatic methods for metaproteomics. His work spans from environmental microbiology (i.e., microbial communities in soil and groundwater ecosystems) to bioenergy (engineering of microbial solubilization of cellulosic biomass to generate biofuels and bioproducts) and finally to the human microbiome (characterizing the microbial connections between human health and disease).

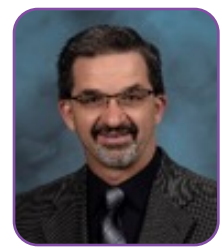

**Yehia Ibrahim** is a senior scientist in the Biological Sciences Division at PNNL. He earned his B.Sc. and M.Sc. degrees in chemistry from Cairo University and his PhD in Chemistry from Virginia Commonwealth University, before joining PNNL as a postdoctoral researcher. Dr. Ibrahim has authored over 110 papers and owns 32 US patents in mass spectrometry and ion mobility spectrometry (IMS). His research focuses on developing ion mobility mass spectrometry-based technologies for enhanced molecular characterization in proteomics and metabolomics. Dr. Ibrahim's work aided the development of a new ultrasensitive IMS-TOF platform for biomolecule analysis for which PNNL was awarded the 2013 R&D 100, currently commercialized by Agilent Technologies (6560 IMS-QTOF). Currently, he is leading research efforts for the development of Structures for Lossless Ion Manipulations (SLIM) technology, also awarded the 2017 R&D 100. Furthermore, his team was awarded the 2018 Federal Laboratory Consortium Award for Excellence in Technology Transfer of SLIM technology to MOBILion Systems Inc. In 2019, Dr. Ibrahim was awarded the PNNL Inventor of the Year and the Distinguished Inventor of Battelle. Dr. Ibrahim has also published on thermochemistry, kinetics, and laser spectroscopy of ion molecules reactions and clusters structures using ion mobility and molecular orbital calculations.

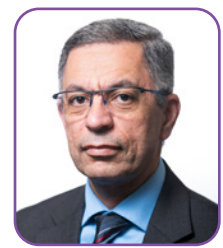

**Tom Metz** is a Laboratory Fellow and Principal Investigator within the Integrative Omics Group at Pacific Northwest National Laboratory (PNNL). He is a bioanalytical chemist by training and received a PhD in chemistry from the University of South Carolina, followed by post-doctoral studies in mass spectrometry at PNNL. His research has focused primarily on applying mass spectrometry-based omics approaches, including proteomics, in studies of diabetes mellitus and infectious diseases, resulting in over 180 publications to date. More recently, he has pioneered the exploration of a "reference-free" compound identification paradigm at PNNL through the implementation of multi-dimensional analytical measurements coupled with computational predictions of measured properties. Currently, Dr. Metz is PI of the Pacific Northwest Advanced Compound Identification Core within the NIH Common Fund Metabolomics Program, PI of the Proteomics Laboratory for The Environmental Determinants of Diabetes in the Young consortium, Lead of the PNNL *m/q* Initiative, and President of the Metabolomics Association of North America.

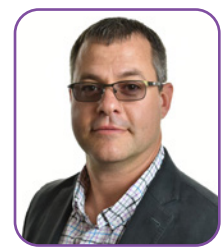

**Trent Northen** is Deputy Division Director and a Senior Scientist within the Environmental Genomics and Systems Biology Division at Berkeley Lab, Adjunct Professor of Comparative Biochemistry at UC Berkeley, the Metabolomics Program Lead at the DOE Joint Genome Institute, Laboratory Research Manager for the multi-institutional m-CAFEs SFA program, and Director of High Throughput Biochemistry at the Joint Genome Institute. Dr. Northen obtained his BS in Chemical Engineering at the University of California Santa Barbara, his PhD in Chemistry and Biochemistry from Arizona State University and performed Post-Doctoral Fellow at the Scripps Research Institute. He has received numerous awards including a DOE Early Career Award, two R&D100 awards, Berkeley Lab Inventor of the Year and was awarded a Presidential Award for Science and Engineering (PECASE) by President Obama. Dr. Northen's laboratory focuses on understanding the role of exogenous small molecule metabolites in mediating microbial interactions with other microbes and plant hosts and how these processes impact soil carbon cycling. Dr. Northen has also championed the development of fabricated ecosystems spanning scales and complexity. Most recently he has led a project at Berkeley Lab developing the 'EcoBOT' which automates plant-microbe-environment studies and integrates lab and field studies. http://www.northenlab.org/

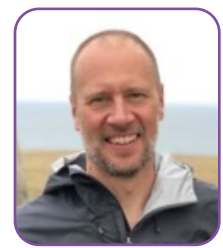

**Simone Raugei** is a chief scientist at PNNL, where his research focuses on developing and applying computational and theoretical methodologies for studying chemical and biochemical processes using high-performance computing. He is the PI of the Basic Energy Sciences Biophysical Sciences Program at PNNL that aims at uniquely characterizing key biochemical and biophysical features of enzymatic processes related to the production of a suite of small sustainable energy carriers to drive the design of synthetic catalytic platforms with enhanced performances. Dr. Raugei is also the PI for the crosscutting theory thrust for both the Center for Molecular Electrocatalysis, an Energy Frontier Research Center, and the Integrated Institute for Catalysis, where he leads theoretical efforts for the design of catalysts for energy storage and energy delivery based on inexpensive transition metals. Dr. Raugei is the lead for the Molecular Modeling Focus Area of the *m/q* Laboratory Initiative. As a Research Professor at the Institute for Biological Chemistry at Washington State University, he conducts research on the fundamental principles adopted by redox enzymes that provide molecular-level control of the catalytic bias—the capability of certain reversible enzymes to accelerate a reaction in one direction significantly different than the reverse direction.

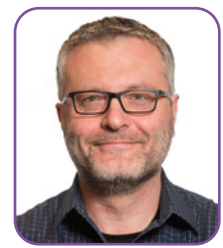

**Rachel Richardson** is a data scientist for the Data Science and Biostatistics team at Pacific Northwest National Laboratory. She specializes in bioinformatics projects utilizing 'omics data. Her previous professional work includes working as a QC analyst for Seattle Genetics and in a microbial research lab during her undergraduate studies, both providing insights into why datasets behave in particular ways. Rachel's current work strongly relies on workflow development, building skills in overall programming as well as application and R package development. Current projects include development of a GUI to assist users in selecting an appropriate machine learning algorithm for uploaded 'omics data and investigating the performance of internal standard based normalizations compared to other methods for lipidomic data.

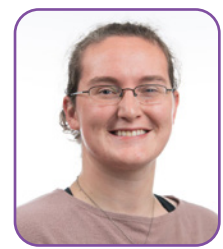

**Paul Rigor** is a Cloud Architect with the Center for Cloud Computing at PNNL. Dr. Rigor has a PhD in Computer Science with over 18 years of experience designing, developing, and delivering technologies and products on traditional high-performance computing and cloud computing environments. Most recently, he was a Data Scientist and AI/ML Strategist with Amazon Web Services, where he led and advised teams from both large Fortune 200 companies as well as growing startups—from ideation to productization. As a former Research Scientist with Verizon, he also holds patents related to storage optimizations, network security, and network performance for global content delivery networks. His academic research interests include the intersection of machine learning, molecular dynamics, genomics, and bioinformatics for drug repositioning.

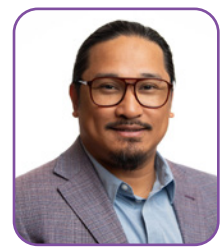

**Kabrena Rodda** is a group lead at PNNL and leads a capability strategy on improving U.S. chemical threat recognition and response. Dr. Rodda is a committee member for the National Academy of Science, Engineering, and Medicine (NASEM) Study, "Assessing and Improving Strategies for Preventing, Countering, and Responding to Weapons of Mass Destruction: Chemical Threats" and represented NASEM as a domestic expert in support of the Government Accountability Office's Technology Assessment on the Forensic Attribution of Chemical Weapons. During her 22-year career as an Air Force scientist, she managed a $30M non-proliferation program and advised on chemical issues at the National Counterproliferation and Biosecurity Center, before retiring as a Colonel. She was a UN Special Commission inspector and lab chief in Iraq in 1995 and 1998 and provided consequence management advice for the Sydney Olympics. In 2017 and 2018, she led chemical threat response workshops at OPCW and headed the writing team for the American Chemical Society (ACS) policy statement, "Preventing the Reemergence of Chemical Weapons." Dr. Rodda is a recipient of the OPCW Director General's Medal, the Secretary of Energy Appreciation Award, and the Secretary of the USAF R&D Award. She holds a PhD in forensic toxicology.

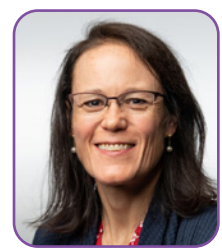

**Richard Smith** is a Battelle Fellow and Chief Scientist in the Biological Sciences Division at Pacific Northwest National Laboratory (PNNL). Dr. Smith has presented more than 300 invited or plenary lectures at national and international scientific meetings, holds 80 US patents, and is the author or co-author of more than 1100 peer reviewed scientific publications. His research has included the development of separations based upon supercritical fluid chromatography, capillary electrophoresis, capillary LC, and Structures for Lossless Ion Manipulations high resolution ion mobility spectrometry in conjunction with Ultra-high performance mass spectrometry. His honors include being the recipient of the 2003 American Chemical Society Award for Analytical Chemistry, the 2009 HUPO Discovery Award in Proteomics Sciences, and the 2013 ASMS Distinguished Contribution in Mass Spectrometry Award. Dr. Smith's research interests include the development and applications of much more effective and informative capabilities, platforms and informatic approaches for proteomics, metabolomics, and other 'pan-omics' measurements in environmental and biological research.

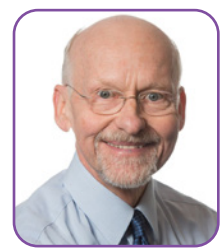

**Justin Teeguarden** is a Laboratory Fellow and the Chief Science Officer for the Department of Energy's Environmental Molecular Sciences Laboratory (EMSL) at the Pacific Northwest National Laboratory where he leads development and execution of EMSL's long-term science and technology strategy. Previously, Dr. Teeguarden led the Decoding the Molecular Universe Directorate Objective for the Earth and Biological Sciences Directorate of PNNL and the Defense Health Programs for the National Security Directorate. His professional experience is in computational and experimental exposure assessment in humans, animals, and cell culture systems. His particular focus has been the utilization of emerging technologies, novel experimental data, and computational methods for addressing public health challenges related to human exposure to chemicals. His experience includes developing pharmacokinetic models of chemicals and particles as tools for understanding the relationship between external exposure and internal exposures in humans and test systems. Dr. Teeguarden has served on several National Academy of Sciences Committees related to exposure science, toxicology and risk assessment. He currently serves on the U.S. EPA's Board of Scientific Counselors and is a member of Board of Directors for the Health and Environmental Sciences Institute. Dr. Teeguarden received a PhD in toxicology from the University of Wisconsin, Madison and is an Eagle Scout.

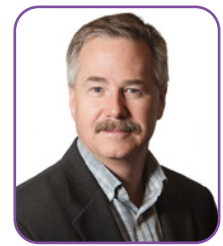

**Bobbie-Jo Webb-Robertson** is the Director of the Biological Sciences Division with the Earth and Biological Sciences Directorate at Pacific Northwest National Laboratory (PNNL). Dr. Webb-Robertson is a statistician by training and received a PhD in decision sciences and engineering systems from Rensselaer Polytechnic Institute. Her research focuses on the development of machine learning (ML) and statistical methods in two primary areas; improving downstream analytics from mass spectrometry derived proteomic, metabolomic and lipidomic data and machine learning driven feature extraction focused on biomarker discovery from complex heterogenous data. Currently, she is a PI leading the development of Artificial Intelligence (AI)/ML-ready data for studies of diabetes mellitus funded through the Human Islet Research Network from the National Institutes of Health, co-I across an array of biomedical projects, as well as the lead for the statistics and machine learning focus area of the PNNL *m/q* Initiative.

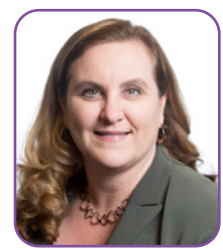

**Katrina Waters** is a Laboratory Fellow and Chief Scientist in the Earth & Biological Sciences Directorate at PNNL. Dr. Waters is internationally known for leadership and innovation in systems biology and bioinformatics to impact global human health challenges at the intersection of environmental exposures and infectious disease. Dr. Waters currently leads the Predictive Phenomics Science and Technology Initiative at PNNL and is the Chair of the Council of Fellows. She recently led a Department of Energy research program focused on airborne and environmental transmission of COVID-19. She has also led numerous research efforts in Computational Modeling, Bioinformatics, and Data Management for a NIAID Center for Predictive Modeling of Infectious Diseases and a Department of Homeland Security program for Predictive Modeling of Viral Infections. Dr. Waters holds joint faculty appointments with the University of Washington and Oregon State University. She has a PhD in Biochemistry from the University of Wisconsin at Madison.

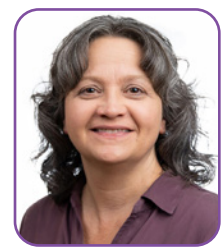

# ACADEMIA

**Peter Armentrout,** PhD, is a Professor of Physical and Analytical Chemistry at the University of Utah. Research interests include thermochemistry, kinetics, and dynamics of simple and complex chemical reactions involving transition metals, heavy elements, and biological molecules. Dr. Armentrout was recognized as a Distinguished Professor of Chemistry and Cannon Fellow in 2003 and is presently the Henry Eyring Presidential Endowed Chair in Chemistry. Dr. Armentrout has been a member of the editorial advisory boards for several journals, has more than 560 research publications that have appeared in the literature, has received several awards for his achievements in Mass Spectrometry, and is a Fellow of several Societies. Dr. Armentrout has also been recognized for his teaching skills, being awarded the R.W. Parry Teaching Award by the Chemistry Department at the University of Utah. More than 40 students have received their PhD's under Professor Armentrout's tutelage.

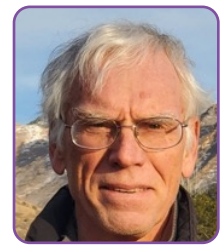

**Charles DeLisi** is Metcalf Professor of Science and Engineering at Boston University, and Dean Emeritus of the College of Engineering. Dr. DeLisi has authored or coauthored more than 300 research papers in various areas of science, including molecular genetics, physical chemistry, applied mathematics and bioinformatics, the last of these being a field in which he played a prominent role as founder. He has given hundreds of invited lectures at national and international conferences on every continent, has mentored more than 150 students, and has collaborated with colleagues world-wide from numerous disciplines. Dr. DeLisi has served on and chaired dozens of Federal, industrial and university advisory boards and is recipient of numerous honors and awards including the Smithsonian Institution Platinum Technology Award for Pioneering Leadership (shared); the US DOE Bicentennial Exceptional Service Award (Secretary Richardson) and the Informa Clinical and Research Excellence Lifetime Achievement Award. He was sole recipient of the Presidential Citizen's Medal for his role in initiating the Human Genome Project. In conferring the Medal President Clinton described him as "…scientist, entrepreneur and teacher… in the truest sense, a humanitarian, a man whose life work has been life itself."

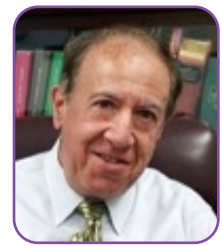

**Katherine (Kate) Duncan** is an Associate Professor in Microbial Antibiotic Discovery at the University of Strathclyde in Glasgow, Scotland. Most of our effective antibiotics are natural products produced by bacteria. However, we are rapidly exhausting our search using traditional methods. It is widely accepted that 'omics methodologies have accelerated our understanding of life at molecular, cellular and organism levels. Our research combines 'omics methods to link biology (genes) to chemistry (antibiotics). We apply this approach to understand the 'chemical language' of microorganisms, and what influences it. This assessment of chemical space across biological parameters can enable informed biodiscovery of new antibiotics. You can find out more about the Duncan lab at: www.medicinesfromthesea.com. She was appointed to a highly competitive Tenure-Track Chancellor's Fellowship at Strathclyde in 2016. Her Fellowship was very productive, and she was put forward for tenure a year early, being awarded full tenure as a Senior Lecturer in 2020. Prior to starting her group, Dr. Duncan completed two Postdoctoral Fellowships at Scripps Institution of Oceanography (California) in Marine Biomedicine and at The Scottish Marine Institute in Marine Biotechnology, a PhD in Biomedical Science (Canada) and a 5-year Masters in Chemistry (Aberdeen, Scotland) with International Placement (Florida).

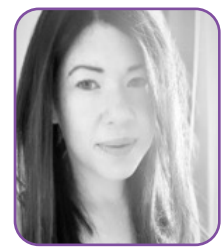

**Facundo Fernández** is a Regents' Professor and Vasser-Woolley Chair in Bioanalytical Chemistry in the School of Chemistry and Biochemistry at the Georgia Institute of Technology, where he also holds the position of Associate Chair for Research and Graduate Training. He received his MSc in Chemistry from the College of Exact and Natural Sciences, Buenos Aires University in 1995 and his PhD in Analytical Chemistry from the same University, in 1999. Between 2000 and 2001, he was a postdoc with Richard N. Zare in the Department of Chemistry at Stanford University. Between 2002–2003, he joined the group of Vicki Wysocki in the Department of Chemistry at the University of Arizona as a senior postdoc and then research scientist. Prof. Fernández is internationally renowned for his work in bioanalytical chemistry, with his research focusing on the development of new tools for assaying small volume samples, tissues and single cells, and applying such methods to better understanding diseases such as cancer, CF and IBD. He is the author of 200+ peer-reviewed publications. He serves on the editorial board of The Analyst and as an Associate editor for the *Journal of the American Society for Mass Spectrometry and Frontiers in Chemistry*.

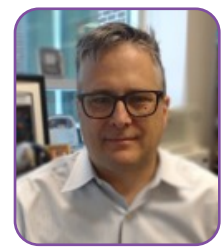

**Oliver Fiehn** has pioneered developments and applications in metabolomics with over 420 publications, starting in 1998 as postdoctoral scholar and from 2000 onwards as group leader at the Max-Planck Institute in Potsdam, Germany. Since 2004 he is Professor at the University of California, Davis, Genome Center, overseeing his research laboratory and the satellite core service laboratory in metabolomics research. Since 2012, he serves as Director of the UC Davis West Coast Metabolomics Center, supervising 35 staff operating 17 mass spectrometers and coordinating activities with four UC Davis satellite labs. Professor Fiehn's research aims at understanding metabolism on a comprehensive level. To leverage data from diverse sets of biological systems, his research laboratory focuses on standardizing metabolomic reports and establishing metabolomic databases and libraries (e.g., MassBank of North America that hosts over 2m public metabolite mass spectra http://massbank.us). Professor Fiehn's laboratory members develop new approaches in analytical chemistry for covering the metabolome, from increasing peak capacity by ion mobility to compound identifications through cheminformatics workflows and software. He applies metabolomics to metabolic questions in human population cohorts, animal and plant models, cells and microorganisms. He also studies fundamental biochemical questions from metabolite damage repair to the new concept of epimetabolites.

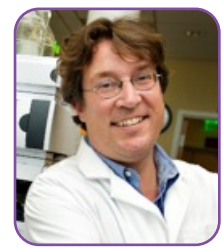

**Augustus Way Fountain III** teaches chemistry and forensics at the University of South Carolina. Prior to teaching at USC, Dr. Fountain had 35 years of military and civilian service with the U.S. Army. As a member of the Scientific and Professional (ST) cadre of the Senior Executive Service he served as the Senior Research Scientist for Chemistry and Acting Director of the Research and Technology Directorate at the Edgewood Chemical Biological Center, and the U.S. Army Deputy Chief Scientist at ASA(ALT). He is an internationally recognized expert in chemical, biological, radiological, nuclear and explosives (CBRNE) sensing. He served as an at-large representative of the United States to NATO for CBRNE and chaired the NATO Sensors & Electronics Technology Panel. As a military officer, Dr. Fountain served with the 1$^{st}$ Ranger Battalion in Operation Just Cause, the 82$^{nd}$ Airborne Division during Operations Desert Shield and Desert Storm, and as a Professor of Chemistry at the United States Military Academy. In 2016, Dr. Fountain was honored with a Meritorious Senior Professional Presidential Rank Award and in 2013 he was awarded the Department of the Army Research & Development Achievement Award for Technical Excellence. He is an elected Fellow of SPIE.

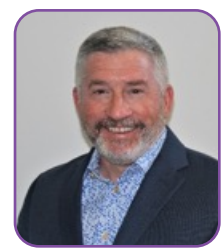

**Neha Garg** is an Assistant Professor at Georgia Tech broadly interested in understanding how small molecules shape microbial composition in complex environments. Dr. Garg obtained her PhD in 2013 from the University of Illinois at Urbana-Champaign with Professors Wilfred A. van der Donk and Satish K. Nair. Dr. Garg's dissertation work in Illinois was recognized by the Anne A Johnson work award and the Catherine Connor Outstanding Dissertation in Biotechnology award. She then worked with Professor Pieter C. Dorrestein as a postdoctoral research associate at the University of California, San Diego where she developed metabolomics methods to visualize microbial communities. Dr. Garg was awarded NSF CAREER Award, NIH R35 MIRA award, and Sandia's Laboratory Directed Research and Development award to develop -omics methods for biomarker discovery and for investigating the function and regulation of microbially-produced natural products with the long-term goal of identifying probiotic candidates for treatment of diseases in human and marine corals. The Garg laboratory applies interdisciplinary approaches in mass spectrometry, microbiology, microscopy, and genomics to unveil the role of microbial, host, and chemical environments in production of small molecule natural products.

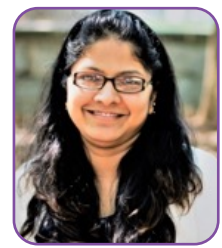

**Stefan Grimme** studied Chemistry and finished his PhD in 1991 in Physical Chemistry on a topic in laser spectroscopy. Dr. Grimme did his habilitation in Theoretical Chemistry in the group of Sigrid Peyerimhoff. In 2000 he got the chair for Theoretical Organic Chemistry at the University of Muenster. In 2011 he accepted an offer as the head of the newly founded Mulliken Center for Theoretical Chemistry at the University of Bonn. He has published more than 630 research articles and is the recipient of the 2013 Schroedinger medal of the World Organization of Theoretically Oriented Chemists (WATOC). In 2014 he was awarded the "Gottfried Wilhelm Leibniz-Preis" from the DFG (endowed by 2.5 million Euro). His main research interests are the development and application of quantum chemical methods for large molecules, density functional theory, non-covalent interactions, conformational analysis and theoretical spectroscopy including mass spectrometry. Currently he is intensively developing next-generation semiempirical QM methods for application in automated computational screening workflows.

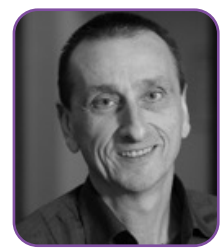

**Roger Linington** is a Professor of Chemistry at Simon Fraser University in Canada where he holds a Tier 1 Canada Research Chair in Natural Products and High-Throughput Screening. His research program focuses developing new tools in 1) chemical characterization of complex mixtures, 2) phenotypic fingerprinting of bioactive metabolites and 3) creation of informatics platforms to integrate chemical and biological datasets. Professor Linington's expertise integrates over 20 years of wet lab science in small molecule characterization with informatics tool development through the creation of open-source databases, webservers and informatics pipelines for small molecule discovery and characterization. Professor Linington brings experience in both the practical aspects of small molecule identification and the technical aspects of developing new computational methods in this area. This experience includes separation science, nuclear magnetic resonance spectroscopy, mass spectrometry, database design, algorithm development. Recent projects from his group include the Natural Products Atlas (www.npatlas.org), a database of all known microbial metabolites, NP Analyst (www.npanalyst.org), a webserver for bioactive compound prediction from complex libraries, and SNAP-MS (www.npatlas.org/discover/snapms), an online tool for reference-free subnetwork annotation of GNPS networks.

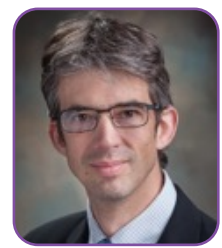

**Gary Miller** serves as Vice Dean for Research Strategy and Innovation and Professor of Environmental Health Sciences in the Mailman School of Public Health, and Professor of Molecular Pharmacology and Therapeutics in the Vagelos College of Physicians and Surgeons at Columbia University in New York. He completed his PhD in Pharmacology and Toxicology and postdoctoral training in molecular neuroscience. His laboratory studies the role of environmental factors in neurodegenerative diseases, including Parkinson's disease and Alzheimer's disease. Dr. Miller founded the first exposome center in the U.S. and wrote the first book on the topic. His team focuses on the use of high-resolution mass spectrometry methods to provide an omic-scale analysis of the human exposome. He is a member of the National Institutes of Health All of Us Research Program Advisory Panel and the National Institute of Environmental Health Sciences Advisory Council. Dr. Miller is the founding editor of the new journal Exposome, published by Oxford University Press.

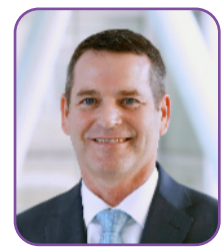

**Jeffrey Nivala** is a Research Assistant Professor in the University of Washington's Allen School of Computer Science. He lead a group focused on molecular technology development, with applications in areas such as data storage, synthetic biology, genomics, and proteomics, particularly interested in nanopore technology for single-molecule protein sequencing. Prior to this, Dr. Nivala was a postdoc at Harvard Medical School where he worked on in-vivo molecular recording technologies using CRISPR.

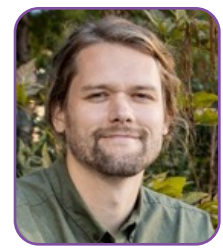

**Aristides Patrinos** is a member of the Novim Group and a Visiting Scholar at the New York University Center for Neural Science. From 2006 to 2020 he worked for Synthetic Genomics Inc, in various roles including as President. Since 2006 he is also a member of the Board of Directors of Tsakos Energy Navigation, a public fuel transportation company listed as TNP in the NYSE. Dr. Patrinos also serves the Advisory Board of EdenRoc Sciences, a privately held biotechnology company; and the Science Advisory Board of DataCubed Inc., a NYC-based private company focused on healthcare, big data, and human decision-making. From 1976 to 2006 he served the U.S. Department of Energy (DOE) and several DOE National Laboratories and led key research programs in biology, medicine, and the environment, including global climate change. He had a leading role in the Human Genome Project and in the "genomics" revolution. He is the recipient of three U.S. Presidential Rank Awards, and two Secretary of Energy Gold Medals. He holds degrees from the National Technical University of Athens and Northwestern University in Evanston, Illinois. His scientific interests include global climate change prediction and biotechnology for climate change mitigation.

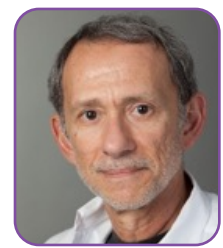

**Dan Raftery** is a Medical Education and Research Endowed Professor at the University of Washington, School of Medicine, and Professor at the Fred Hutchinson Cancer Center in Seattle, WA. Dr. Raftery received his PhD from Berkeley and was previously Professor of Chemistry in the Analytical Division at Purdue University, where his group began its research in metabolomics in 2003. Dr. Raftery is the Founding Director of the Northwest Metabolomics Research Center at UW Medicine, and an Associate Editor for Analytical Chemistry. Dr. Raftery's current research program is focused on the development of new analytical methods in metabolomics and their application to a range of clinical and basic science studies related to metabolic alterations. His group uses advanced mass spectrometry and NMR methods for the identification of early biomarkers and metabolic risk factors for numerous cancers and other diseases, and for the exploration of systems biology in cells and mitochondria.

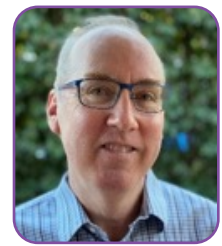

**Akos Vertes** is a Professor of Chemistry at the George Washington University in Washington, DC. His research interests encompass the development of new analytical techniques applicable in diverse fields of chemistry, biology, and medicine. Research areas include single cell and subcellular analysis, high throughput and ultrasensitive methods applicable in systems biology, proteomics and metabolomics, and new methods for molecular imaging of biological tissues. His research has been presented in over 200 peer-reviewed publications ($h$ = 51), and in two books. Dr. Vertes is a co-inventor on 19 patents. His patents have been licensed and commercialized by multiple industrial partners. He was elected Fellow of the American Association for the Advancement of Science, Fellow of the National Academy of Inventors, and received the Distinguished Researcher Award at GWU, the 2012 Hillebrand Prize and the Oscar and Shoshana Trachtenberg Prize for Scholarship. He is a Doctor of the Hungarian Academy of Sciences. He served as Visiting Faculty at the Lawrence Berkeley National Laboratory, an MTA Distinguished Guest Scientist at the Hungarian Academy of Sciences in Hungary, and twice as a Visiting Professor at the Swiss Federal Institute of Technology Zurich (ETH Zurich) in Switzerland.

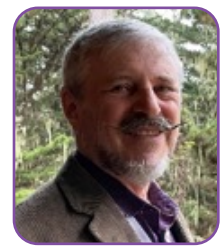

**Nelson Vinueza** is an Associate Professor and Faculty Scholar in the Department of Textile Engineering, Chemistry, and Science and the Department of Chemistry at North Carolina State University. He joined NC State in August 2013 as part of the Chancellor's Faculty Excellence Program. His research centers around three aspects of mass spectrometry: fundamentals, instrumentation, and applications in dyes, textiles, forensics, and biofuels. Dr. Vinueza is the Max Weaver Dye Library Director, a historical collection with approximately 100,000 synthetic dyes, where he has been creating a database since 2016 to enhance the use of these organic molecules in different fields through machine learning. This work is currently being used for several federal and industrial research projects. He has over 12 years of experience in small molecule characterization by developing new mass spectrometric methods. In 2022 he received the Dyers' Company Research Medal. Dr. Vinueza earned his PhD in Physical Organic Chemistry from Purdue University, where he studied the chemical reactivity of carbon-center tri- and tetraradicals, and later conducted post-doctoral studies at the Center of Direct Catalytic Conversion of Biomass to Biofuels at Purdue University where he developed new mass spectrometry methods for analyzing lignin and cellulose degradation products and bio-oil.

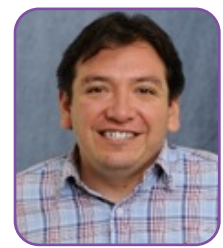

**David Wishart** (PhD Yale, 1991) is a Distinguished University Professor in the Departments of Biological Sciences and Computing Science at the University of Alberta. He also holds adjunct appointments with the Faculty of Pharmaceutical Sciences and with the Department of Pathology and Laboratory Medicine. Dr. Wishart's research interests include metabolomics, analytical chemistry, food chemistry, natural product chemistry, molecular biology, protein chemistry and neuroscience. He has developed a number of widely used techniques based on NMR spectroscopy, mass spectrometry, liquid chromatography and gas chromatography to characterize the structures of both small and large molecules. As part of this effort, he has led the "Human Metabolome Project" (HMP), a multi-university, multi-investigator project that is cataloguing all known chemicals in human tissues and biofluids, which has identified or found evidence for more than 250,000 metabolites in the human body, archived on the Human Metabolome Database (HMDB). His lab has also been using machine learning and artificial intelligence techniques to help create other useful chemistry databases and software tools to help with the characterization and identification of metabolites, drugs, pesticides and natural products. Over the course of his career Dr. Wishart has published more than 500 research papers.

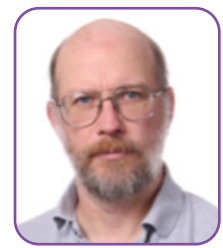

# GOVERNMENT AGENCIES

**David Balshaw** is the Director of the Division of Extramural Research and Training at the National Institute for Environmental Health Sciences. He and his colleagues oversee NIEHS's grants programs, with an annual investment of over $500M in research spanning from fundamental technology development to mechanistic research to community engaged intervention and dissemination research. Dr. Balshaw is a biophysicist by training with a PhD from the Department of Pharmacology and Cellular Biophysics at the University of Cincinnati and post-doctoral training in the Department of Biochemistry and Biophysics at the University of North Carolina. He joined NIH as a Program Officer at the National Heart, Lung, and Blood Institute in 2001 and moved to NIEHS in 2003. Over nearly two decades he has led several high impact programs at NIEHS and NIH particularly in the areas of 'omics technologies and is an international leader in the field of exposomics.

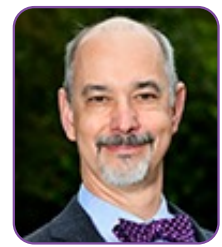

**Evan Bolton** is Program Head of Chemistry at the National Center for Biotechnology Information and is in charge of PubChem (https://pubchem.ncbi.nlm.nih.gov), among other things. PubChem is an open repository for chemical substance information and their bioactivities used by millions of users a month. Dr. Bolton (https://www.linkedin.com/in/pubchem) has been with the PubChem project since its inception (nearly 20 years). With over sixty publications (https://scholar.google.com/citations?user=8d-zWEMAAAAJ), many highly cited, Dr. Bolton has been involved in a broad array of community-based efforts, in addition to PubChem and its two subprojects: PubChem3D and PubChemRDF.

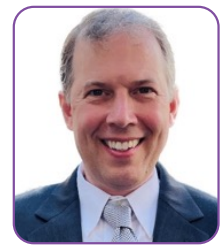

**Donald Cronce** is the Threat Agent Science Lead at the US Defense Threat Reduction Agency, Research and Development Directorate, Chemical and Biological Technologies Department and the Advanced and Emerging Threat Division. He has a Bachelor of Science degree in Chemistry from the University of Wisconsin at Milwaukee and a PhD in Chemistry from The Pennsylvania State University (University Park, PA). Dr. Cronce has authored numerous technical presentations and articles, and holds five US patents in CBRN defense-related areas. He has conducted, managed and/or directed R&D in many areas of CBRN defense for over 30 years, from basic research to advanced development.

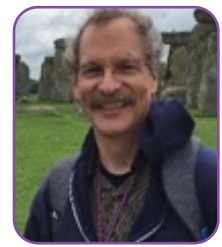

**Dan Drell** grew up on the campus of Stanford University; he majored in biology at Harvard, got his PhD in immunology at the University of Alberta in Edmonton, Alberta, Canada, and went on to a career in the sciences, with stops for research at the Sloan-Kettering and Rockefeller University in New York, and Baylor College of Medicine before moving to the WDC area. There he worked at NIH for 3 years (diabetes research), George Washington University's Department of Medicine (cancer and tissue typing) and finally, for 27 $\frac{1}{2}$ years, at the Department of Energy's Biological and Environmental Research Office on various activities including the Human Genome Project, Microbial Genome Project, finishing his career as the Program Manager for the Joint Genome Institute User Facility now at LBNL. Since retiring from DOE in 2018, Dr. Drell has joined with several other former DOE officials to explore the uses of synthetic biology for enhanced C02 capture and conversions to value-added products as a step towards addressing climate challenges.

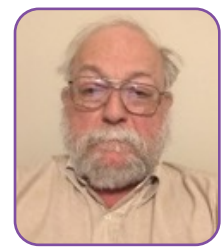

**Christine Fisher (O'Donnell)** is a Chemist in the Center for Food Safety and Applied Nutrition (CFSAN) at the U.S. Food and Drug Administration (FDA) in College Park, MD. She obtained her B.S. degree from Ohio University in 2010 and her PhD from Purdue University in 2015 under the direction of Professor Scott A. McLuckey. In 2015 she joined Merck where she worked as a Senior Scientist in Analytical Research and Development. She began at the FDA as a post-doctoral fellow in 2017 under the mentorship of Dr. Ann Knolhoff and was hired in 2018. At the FDA, Dr. Fisher's research has focused on developing non-targeted analysis (NTA) methods using liquid chromatography high-resolution mass spectrometry for small molecule food safety applications. Specifically, she has developed approaches to improve data quality and prioritize chemicals for identification in complex matrices and is working toward a broadly applicable and accessible standard mixture to aid in assessing NTA method quality. Dr. Fisher is a Co-Chair of the Best Practices for Non-Targeted Analysis (BP4NTA) international working group and has enjoyed collaborating with diverse members to develop tools and resources to improve accessibility and reliability of NTA methods across a variety of fields.

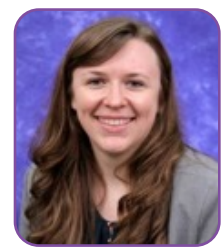

**Trevor Glaros** earned his B.S. in Microbiology from Clemson University (2006) and his PhD in Biology from Virginia Tech (2011). Dr. Glaros did his post-doctoral fellowship at the National Cancer Institute/Frederick National Laboratory using proteomic mass spectrometry to determine the mechanism of action for several leading anticancer pharmaceuticals. Afterwards, he accepted an appointment at the United States Army Medical Research Institute of Infectious Diseases (USAMRIID) where he used imaging mass spectrometry to discover tissue biomarkers of Burkholderia. After his formal training, Dr. Glaros accepted a Principal Investigator role at the U.S. Army Combat Capabilities Development Command Chemical Biological Center (DEVCOM CBC; formerly ECBC) in Edgewood, Maryland. There he founded the laboratory's Mass Spectrometry Core Facility which was widely used for warfighter focused applications. He is internationally recognized as the pioneer for the use of paper spray ionization for the detection of chemical and biological threats. Dr. Glaros joined Los Alamos National Laboratory as the B-11 Deputy Group Leader in 2020 and was promoted to the B-TEK Group leader in 2022. During his time at LANL, Dr. Glaros played a key role in the revitalization of LANL's mass spectrometry capabilities forming the laboratory's first Mass Spectrometry Center for Integrated Omics.

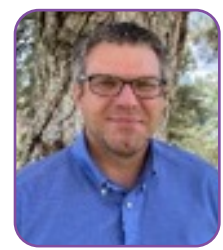

**Rachel Gooding** serves as the Chemical Lead of the Hazard Awareness and Characterization Center in Technology Centers Division within DHS S&T, as well as Chief Scientist in the Office of National Laboratories, Chemical Security Analysis Center. This joint role includes 1) providing technical advice and expertise on chemical threats, hazards, and risks to decision makers and RDT&E programs, 2) conducting research to inform DHS high priority requirements and S&T's R&D program formulation, and 3) establishing and maintaining scientific and technical partnerships in chemical and related disciplines. Current initiatives include computational approaches to predictive toxicology and AI/ML methods for identifying developments in the chemical and biochemical fields relevant to the DHS mission, such as chemical synthesis, target-receptor interactions, and detection, and risk modeling. She has been supporting DHS S&T since 2007. Before joining the government, Dr. Gooding supported the private sector on chemical processing modeling, field trials of low vapor pressure detectors and pilot scale construction and evaluation of systems for the destruction of chemical warfare agents. She has over 29 years of experience in toxic chemical characterization and chemical defense related work. Dr. Gooding has a BS in chemistry and has completed graduate studies in chemical engineering, focusing on catalysis and reaction engineering.

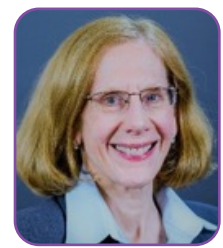

**Rabih Jabbour** holds PhD in Bioanalytical chemistry with more than 20 years of research experience in academia and industry in the field of CBRNE detection and technologies development and evaluations. He developed automated sample processing system for extraction and preconcentration of biological biomarkers for the detection and identification microbes down to strain level. He co-authors three patents in the detection field and contributed to various government projects such as water monitoring, microbial imaging, Organ on a chip, and portable and wearable detection and sensing devices. He has collaboration with government, national and international laboratories and academia. Dr. Jabbour is serving as Laboratory manager for the CSAC Chemistry Security Laboratory and gained leadership and organizational skills through work experience and participation in various leadership, organizational, coaching, and project management courses for the last 10 years.

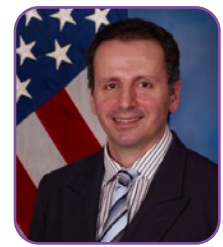

**Conor Jenkin**s received his B.S. in Chemistry from Ithaca College in upstate New York conducting research in computation chemistry. From there he joined the Proteomics Core Facility at the National Cancer institute where he fell in love with multi-omics. After obtaining his master's in bioinformatics at Hood College, he currently is employed by the U.S. Army's DEVCOM CBC Biodefense division where he investigates biological and chemical threat agents by performing untargeted proteomic, metabolomic and lipidomic analyses on biological samples exposed to these agents. By performing network analysis on his data, mechanism of action and potential medicinal countermeasures can be discovered to aid the warfighter in their operation. He is currently obtaining his PhD in Biochemistry at the University of Maryland - College Park researching RNA degradation in microbial organisms.

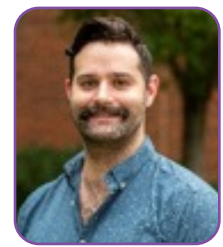

**Ann Knolhoff** is the Branch Chief for the Spectroscopy and Mass Spectrometry Branch (SMSB) at the Food and Drug Administration's (FDA's) Center for Food Safety and Applied Nutrition. She received her B.Sc. in Chemistry at Truman State University and her PhD in Chemistry at the University of Illinois at Urbana-Champaign under the direction of Professor Jonathan Sweedler. Dr. Knolhoff has been at the FDA since 2011, first as a post-doctoral fellow under the direction of Dr. Tim Croley and then transitioning to a Research Chemist in SMSB where her research primarily focused on the development of non-targeted analysis approaches using liquid chromatography and high-resolution mass spectrometry for food safety applications. The goal of this work was to determine best practices for non-targeted analysis which could be applied to a variety of different sample types, capable of detecting and identifying broad molecular species, and yielding reliable and reproducible results. As Branch Chief, she directs SMSB researchers working on diverse applications and projects, including food additives, allergens, protein toxins, plant-incorporated protectants, species identification, nanomaterials, economic adulteration, and non-targeted analysis in CFSAN-regulated food, packaging, and cosmetic products.

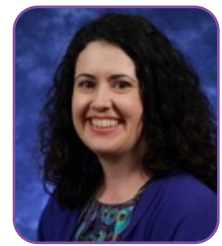

**Morgan Minyard** is the branch chief of Molecular Toxicology at the U.S. Army Combat Capabilities Development Command Chemical and Biological Center (DEVCOM CBC). As branch chief, she oversees the predicative toxicology and ecotoxicity research programs. These programs are to inform the warfighter of the hazard they face by providing early characteristics of chemicals of interest. Prior to her role as branch chief, Dr. Minyard was a team lead and science and technology manager at the Defense Threat Reduction Agency managing portfolios under the advanced and emerging threats division and the detection and diagnostics division. Her educational achievements include a doctorate in soil science and biogeochemistry and a masters in environmental engineering, both awarded by the Pennsylvania State University. She received a bachelor's degree from the University of Colorado in molecular, cellular, and development biology and biochemistry.

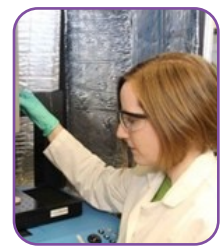

**Kirsten Overdahl** is a mass spectrometrist and environmental health researcher dedicated to exploring emerging environmental contaminants and their health implications through metabolomics and exposomics. Dr. Overdahl obtained a B.A. in Chemistry with Distinction (2015) from St. Olaf College, where she characterized illicit drugs in forensic and environmental matrices. She then received a PhD in Environmental Toxicology (2021) from Duke University, where she characterized azobenzene disperse dyes in the indoor environment. Dr. Overdahl currently serves as a Chemist and as Trans-NIH Metabolomics Coordinator in the new Metabolomics Core Facility at the National Institute of Environmental Health Sciences (NIEHS), where she collaborates with researchers throughout the National Institutes of Health (NIH) to design, carry out, and analyze metabolomics experiments. Her current independent research interests center around best practices for untargeted analysis data interpretability and, more specifically, for identification of previously uncharacterized chemicals. She is a member of American Society for Mass Spectrometry (ASMS), Society of Environmental Toxicology and Chemistry (SETAC), and Best Practices for Non-Targeted Analysis (BP4NTA). Dr. Overdahl currently serves on the Executive Board of Triangle-Area Mass Spectrometry (TAMS), is the co-chair for SETAC Women in Science, and is a member of NIEHS DIR's Diversity, Equity, Inclusion, and Accessibility (DEIA) working group.

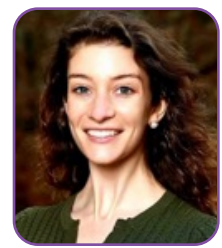

**Douglas Sheeley** is currently the Acting Director of the NIH Office of Strategic Coordination (OSC). This Office is a component of the Division of Program Coordination, Planning, and Strategic Initiatives (DPCPSI) in the NIH Office of the Director and is responsible for management of the NIH Common Fund. Dr. Sheeley joined OSC in 2019 as a program leader and has served as OSC Deputy Director since 2022. As a Program Director and Senior Scientific Officer at the former National Center for Research Resources (NCRR) and NIGMS (2000–2017), he led programs developing new technologies to catalyze scientific advances in structure determination, proteomics and metabolomics, and informatics, including several Common Fund programs. Dr. Sheeley served in multiple leadership roles at NCRR and NIGMS, and as the Deputy Director of NIDCR. Prior to NIH, he was a researcher at Glaxo Wellcome Research & Development, where he gained an industry perspective on project management and the value of diverse multidisciplinary teams. He earned his doctoral degree in nutritional biochemistry from Harvard University and his BS in chemistry from Dickinson College. His primary research experience is as a bioanalytical chemist, with expertise in biomedical mass spectrometry and structural analysis of both proteins and carbohydrates.

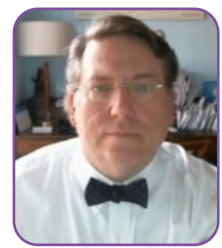

**Jon Sobus** is a senior scientist with the US EPA's Office of Research and Development, Center for Computational Toxicology and Exposure in Research Triangle Park, NC. He received his PhD in Environmental Sciences and Engineering from the University of North Carolina at Chapel Hill in 2008. Jon has worked with EPA/ORD since 2008, having led numerous research teams focusing on human biomonitoring and non-targeted analysis (NTA) research. Since 2016, Dr. Sobus has worked to develop innovative approaches for interpreting high resolution mass spectrometry-based NTA data in an actionable manner. At the core of his efforts, Dr. Sobus aims to develop computational tools that lead to the defensible identification, quantification, and risk-based evaluation of emerging contaminants detected via NTA methods.

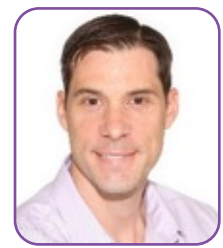

**Marianne Sowa** completed her PhD in physical chemistry and spent her early career working in the areas of ion-molecule interactions, biophysics and reaction kinetics. Dr. Sowa's radiation research career began when she joined the Pacific Northwest National Laboratory in Richland, Washington where she designed and developed a pulsed electron beam single cell irradiator. This device was instrumental in testing the existence of radiation-induced bystander effects for low LET exposures. Dr. Sowa also uses system biology approaches to understanding radiation-induced effects in three-dimensional culture models that more realistically replicate the cellular environment in situ. In 2015, Dr. Sowa joined NASA as Chief of the Space Bioscience Research Branch. In 2018, she became the Chief of the Space Biosciences Division, and currently serves as the Acting Deputy Director for the Science Directorate at NASA, Ames Research Center. In this role, she leads a team of over 150 civil servants and in the Divisions of Earth Science, Space Science and Astrobiology, and Space Biosciences. She is the Center Management lead for their work on the International Space Station and for the Human Research Program. She also served as the Portfolio Scientist for the Space Biology Program where she oversaw both ground and spaceflight experiments.

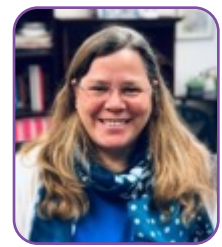

**Antony Williams** is an NMR spectroscopist (PhD London, 1988) and ran the NMR facility at the University of Ottawa and was the NMR Technology Leader at Eastman Kodak. Dr. Williams was the Chief Science Officer for ACD/Labs for over ten years and led the software development for NMR prediction, data processing for NMR, MS, IR and chromatography and Computer-Assisted Structure Elucidation (CASE). While at ACD/Labs he ran a hobby project, ChemSpider, and sold it to the Royal Society of Chemistry. He worked as their Vice President of Strategic Development for over 5 years and joined the Center for Computational Toxicology and Exposure in the Office of Research and Development at US-EPA in May 2015 where he is a cheminformatician focused on the delivery of the center's data to the scientific community. He has over two decades of experience in cheminformatics and chemical information management with a focus on internet-based projects to deliver free-access community-based chemistry websites. He was the product owner for the CompTox Chemicals Dashboard (https://comptox.epa.gov/dashboard) for 7 years and works on multiple projects supporting non-targeted analysis including the development of the DSSTox database, structure standardization, AMOS (the analytical methods and open spectra database) and the EPA/NTA WebApp.

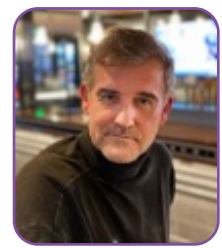

# INDUSTRY

**Chaitanya Gupta** is co-founder and chief technology officer at Probius Inc., a Stanford University spinout that is commercializing Quantum Electrochemical Spectroscopy (QES), an all-electronic transduction paradigm for the sensing of specie vibrations in biological samples. The all-electronic vibration sensing platform leverages machine learning for elucidating analyte or phenotype vibrational signatures from small sample volumes and without prior sample-prep, drastically simplifying biochemical analyses workflows. The Probius platform is being deployed for specimen phenotyping and multi-analyte quantitation for assays in preclinical research settings, with the eventual goal of simplifying disease diagnostics in clinical environments. Dr. Gupta is the inventor of the QES technique, from his PhD thesis work at the University of Illinois at Urbana-Champaign through his postdoctoral appointment at Stanford University. Existing approaches to analysis of biological samples rely on apriori hypotheses to inform the preparatory workflows, biasing the subsequent analyses in a significant manner. Dr. Gupta was inspired by spectroscopy-like approaches that instead create a mathematical representation of the sample from the underlying molecular properties, and which encode the knowns and unknowns in the biochemical space without bias. To that end, QES is a scalable approach to capturing biological information across a broad spectrum of vibrational frequencies.

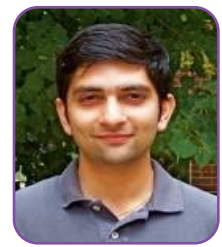

**Navdeep Jaitly** is a Research Scientist at Apple Machine Learning Research where he leads a team of researchers working on a diverse set of fundamental machine learning problems. Dr. Jaitly received his PhD in Computer Science under Geoffrey Hinton at the University of Toronto, during the formative years of Deep Learning. His research interests lie in creating novel machine learning techniques and he has worked on diverse areas such a generative model, self-supervised learning, semi-supervised learning, representation learning, statistical machine learning, etc. On the applied side, he has applied these techniques to diverse areas such as Speech Recognition, Computer Vision, Natural Language Processing, Finance, Computational Biology and Proteomics. In his prior roles he has worked in various research roles in different groups and companies such as Google Brain, Nvidia Research, D. E. Shaw Machine Learning Ventures, PNNL, Caprion Pharmaceuticals and IBM.

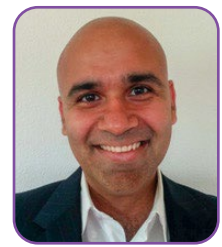

**Tobias Kind** received his PhD in Analytical Chemistry from the University of Leipzig studying chemometrics, gas and liquid chromatography coupled to mass spectrometry for analysis of complex environmental samples in 2003. During his studies Dr. Kind simultaneously worked at the UFZ Centre for Environmental Research Leipzig-Halle, Germany. In 2004 he joined the Metabolomic Analysis group at the Max-Planck-Institute of Molecular Plant Physiology in Potsdam-Golm. The same year, he moved to the newly founded UC Davis Genome Center in California as a postgraduate researcher. For over 18 years Dr. Kind led the cheminformatics team and later on headed the computational core at the NIH West Coast Metabolomics Center for Compound Identification at UC Davis. In 2022 he accepted a position as a Research Fellow at Enveda Biosciences in Boulder Colorado, a drug discovery company that uses machine learning, metabolomics and robotics to discover new drugs from plant natural products. His research projects focus on the advancement of structure elucidation techniques and databases for small molecules, drugs and metabolites. His research interests include machine learning and deep-learning for compound identification, QSAR and ADMET predictions, ion mobility, quantum chemistry, in-silico spectra and ab-initio molecular dynamics for modelling of tandem mass spectra.

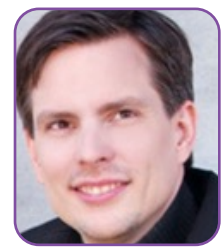

# APPENDIX 4

# REFERENCES

1. Sinsheimer RL. The Santa Cruz Workshop--May 1985. Genomics. 1989;5(4):954-6. doi: 10.1016/0888-7543(89)90142-0.
2. DeLisi C. Meetings that changed the world: Santa Fe 1986: Human genome baby-steps. Nature. 2008;455(7215):876-7. doi: 10.1038/455876a.
3. Dulbecco R. A turning point in cancer research: sequencing the human genome. Science. 1986;231(4742):1055-6. doi: 10.1126/science.3945817.
4. Practice BTP. Economic Impact of the Human Genome Project. 2011 May 2011. Report No.
5. Lakhani CM, Tierney BT, Manrai AK, Yang J, Visscher PM, Patel CJ. Repurposing large health insurance claims data to estimate genetic and environmental contributions in 560 phenotypes. Nat Genet. 2019;51(2):327-34. Epub 20190114. doi: 10.1038/s41588-018-0313-7.
6. Wild CP. Complementing the genome with an "exposome": the outstanding challenge of environmental exposure measurement in molecular epidemiology. Cancer Epidemiol Biomarkers Prev. 2005;14(8):1847-50. doi: 10.1158/1055-9965.EPI-05-0456.
7. Rappaport SM, Smith MT. Epidemiology. Environment and disease risks. Science. 2010;330(6003):460-1. doi: 10.1126/science.1192603.
8. Sud M, Fahy E, Cotter D, Azam K, Vadivelu I, Burant C, Edison A, Fiehn O, Higashi R, Nair KS, Sumner S, Subramaniam S. Metabolomics Workbench: An international repository for metabolomics data and metadata, metabolite standards, protocols, tutorials and training, and analysis tools. Nucleic Acids Res. 2016;44(D1):D463-70. Epub 20151013. doi: 10.1093/nar/gkv1042.
9. Bifarin OO, Gaul DA, Sah S, Arnold RS, Ogan K, Master VA, Roberts DL, Bergquist SH, Petros JA, Edison AS, Fernandez FM. Urine-Based Metabolomics and Machine Learning Reveals Metabolites Associated with Renal Cell Carcinoma Stage. Cancers (Basel). 2021;13(24). Epub 20211213. doi: 10.3390/cancers13246253.
10. Winnike JH, Stewart DA, Pathmasiri WW, McRitchie SL, Sumner SJ. Stable Isotope-Resolved Metabolomic Differences between Hormone-Responsive and Triple-Negative Breast Cancer Cell Lines. Int J Breast Cancer. 2018;2018:2063540. Epub 20180930. doi: 10.1155/2018/2063540.
11. Sun RC, Fan TW, Deng P, Higashi RM, Lane AN, Le AT, Scott TL, Sun Q, Warmoes MO, Yang Y. Noninvasive liquid diet delivery of stable isotopes into mouse models for deep metabolic network tracing. Nat Commun. 2017;8(1):1646. Epub 20171121. doi: 10.1038/s41467-017-01518-z.
12. Chen C, Gowda GAN, Zhu J, Deng L, Gu H, Chiorean EG, Zaid MA, Harrison M, Zhang D, Zhang M, Raftery D. Altered metabolite levels and correlations in patients with colorectal cancer and polyps detected using seemingly unrelated regression analysis. Metabolomics. 2017;13(11). Epub 20170915. doi: 10.1007/s11306-017-1265-0.
13. Sousa CM, Biancur DE, Wang X, Halbrook CJ, Sherman MH, Zhang L, Kremer D, Hwang RF, Witkiewicz AK, Ying H, Asara JM, Evans RM, Cantley LC, Lyssiotis CA, Kimmelman AC. Pancreatic stellate cells support tumour metabolism through autophagic alanine secretion. Nature. 2016;536(7617):479-83. Epub 20160810. doi: 10.1038/nature19084.
14. Budczies J, Pfitzner BM, Gyorffy B, Winzer KJ, Radke C, Dietel M, Fiehn O, Denkert C. Glutamate enrichment as new diagnostic opportunity in breast cancer. Int J Cancer. 2015;136(7):1619-28. Epub 20140904. doi: 10.1002/ijc.29152.
15. Pauling L. Orthomolecular psychiatry. Varying the concentrations of substances normally present in the human body may control mental disease. Science. 1968;160(3825):265-71. doi: 10.1126/science.160.3825.265.

16. Robinson AB, Pauling L. Techniques of orthomolecular diagnosis. Clin Chem. 1974;20(8):961-5.
17. Aharoni A, Goodacre R, Fernie AR. Plant and microbial sciences as key drivers in the development of metabolomics research. Proc Natl Acad Sci U S A. 2023;120(12):e2217383120. Epub 20230317. doi: 10.1073/pnas.2217383120.
18. Hu QZ, Noll RJ, Li HY, Makarov A, Hardman M, Cooks RG. The Orbitrap: a new mass spectrometer. J Mass Spectrom. 2005;40(4):430-43. doi: 10.1002/jms.856.
19. Wang M, Carver JJ, Phelan VV, Sanchez LM, Garg N, Peng Y, Nguyen DD, Watrous J, Kapono CA, Luzzatto-Knaan T, Porto C, Bouslimani A, Melnik AV, Meehan MJ, Liu WT, Crusemann M, Boudreau PD, Esquenazi E, Sandoval-Calderon M, Kersten RD, Pace LA, Quinn RA, Duncan KR, Hsu CC, Floros DJ, Gavilan RG, Kleigrewe K, Northen T, Dutton RJ, Parrot D, Carlson EE, Aigle B, Michelsen CF, Jelsbak L, Sohlenkamp C, Pevzner P, Edlund A, McLean J, Piel J, Murphy BT, Gerwick L, Liaw CC, Yang YL, Humpf HU, Maansson M, Keyzers RA, Sims AC, Johnson AR, Sidebottom AM, Sedio BE, Klitgaard A, Larson CB, P CAB, Torres-Mendoza D, Gonzalez DJ, Silva DB, Marques LM, Demarque DP, Pociute E, O'Neill EC, Briand E, Helfrich EJN, Granatosky EA, Glukhov E, Ryffel F, Houson H, Mohimani H, Kharbush JJ, Zeng Y, Vorholt JA, Kurita KL, Charusanti P, McPhail KL, Nielsen KF, Vuong L, Elfeki M, Traxler MF, Engene N, Koyama N, Vining OB, Baric R, Silva RR, Mascuch SJ, Tomasi S, Jenkins S, Macherla V, Hoffman T, Agarwal V, Williams PG, Dai J, Neupane R, Gurr J, Rodriguez AMC, Lamsa A, Zhang C, Dorrestein K, Duggan BM, Almaliti J, Allard PM, Phapale P, Nothias LF, Alexandrov T, Litaudon M, Wolfender JL, Kyle JE, Metz TO, Peryea T, Nguyen DT, VanLeer D, Shinn P, Jadhav A, Muller R, Waters KM, Shi W, Liu X, Zhang L, Knight R, Jensen PR, Palsson BO, Pogliano K, Linington RG, Gutierrez M, Lopes NP, Gerwick WH, Moore BS, Dorrestein PC, Bandeira N. Sharing and community curation of mass spectrometry data with Global Natural Products Social Molecular Networking. Nat Biotechnol. 2016;34(8):828-37. doi: 10.1038/nbt.3597.
20. Wishart DS, Tzur D, Knox C, Eisner R, Guo AC, Young N, Cheng D, Jewell K, Arndt D, Sawhney S, Fung C, Nikolai L, Lewis M, Coutouly MA, Forsythe I, Tang P, Shrivastava S, Jeroncic K, Stothard P, Amegbey G, Block D, Hau DD, Wagner J, Miniaci J, Clements M, Gebremedhin M, Guo N, Zhang Y, Duggan GE, Macinnis GD, Weljie AM, Dowlatabadi R, Bamforth F, Clive D, Greiner R, Li L, Marrie T, Sykes BD, Vogel HJ, Querengesser L. HMDB: the Human Metabolome Database. Nucleic Acids Res. 2007;35(Database issue):D521-6. doi: 10.1093/nar/gkl923.
21. van der Hooft JJ, Wandy J, Barrett MP, Burgess KE, Rogers S. Topic modeling for untargeted substructure exploration in metabolomics. Proc Natl Acad Sci U S A. 2016;113(48):13738-43. Epub 20161116. doi: 10.1073/pnas.1608041113.
22. Duhrkop K, Nothias LF, Fleischauer M, Reher R, Ludwig M, Hoffmann MA, Petras D, Gerwick WH, Rousu J, Dorrestein PC, Bocker S. Systematic classification of unknown metabolites using high-resolution fragmentation mass spectra. Nat Biotechnol. 2021;39(4):462-71. Epub 20201123. doi: 10.1038/s41587- 020-0740-8.
23. Jarmusch AK, Aron AT, Petras D, Phelan VV, Bittremieux W, Acharya DD, Ahmed MMA, Bauermeister A, Bertin MJ, Boudreau PD, Borges RM, Bowen BP, Brown CJ, Chagas FO, Clevenger KD, Correia MSP, Crandall WJ, Crüsemann M, Damiani T, Fiehn O, Garg N, Gerwick WH, Gilbert JR, Globisch D, Gomes PWP, Heuckeroth S, James CA, Jarmusch SA, Kakhkhorov SA, Kang KB, Kersten RD, Kim H, Kirk RD, Kohlbacher O, Kontou EE, Liu K, Lizama-Chamu I, Luu GT, Knaan TL, Marty MT, McAvoy AC, McCall L- I, Mohamed OG, Nahor O, Niedermeyer THJ, Northen TR, Overdahl KE, Pluskal T, Rainer J, Reher R, Rodriguez E,

Sachsenberg TT, Sanchez LM, Schmid R, Stevens C, Tian Z, Tripathi A, Tsugawa H, Nishida K, Matsuzawa Y, Hooft JJJvd, Vicini A, Walter A, Weber T, Xiong Q, Xu T, Zhao HN, Dorrestein PC, Wang M. A Universal Language for Finding Mass Spectrometry Data Patterns. bioRxiv. 2022:2022.08.06.503000. doi: 10.1101/2022.08.06.503000.

24. Wang M, Jarmusch AK, Vargas F, Aksenov AA, Gauglitz JM, Weldon K, Petras D, da Silva R, Quinn R, Melnik AV, van der Hooft JJJ, Caraballo-Rodriguez AM, Nothias LF, Aceves CM, Panitchpakdi M, Brown E, Di Ottavio F, Sikora N, Elijah EO, Labarta-Bajo L, Gentry EC, Shalapour S, Kyle KE, Puckett SP, Watrous JD, Carpenter CS, Bouslimani A, Ernst M, Swafford AD, Zuniga EI, Balunas MJ, Klassen JL, Loomba R, Knight R, Bandeira N, Dorrestein PC. Mass spectrometry searches using MASST. Nat Biotechnol. 2020;38(1):23-6. doi: 10.1038/s41587-019-0375-9.
25. Kim S, Thiessen PA, Bolton EE, Chen J, Fu G, Gindulyte A, Han L, He J, He S, Shoemaker BA, Wang J, Yu B, Zhang J, Bryant SH. PubChem Substance and Compound databases. Nucleic Acids Res. 2016;44(D1):D1202-13. Epub 20150922. doi: 10.1093/nar/gkv951.
26. Grulke CM, Williams AJ, Thillanadarajah I, Richard AM. EPA's DSSTox database: History of development of a curated chemistry resource supporting computational toxicology research. Comput Toxicol. 2019;12. doi: 10.1016/j.comtox.2019.100096.
27. Sorokina M, Merseburger P, Rajan K, Yirik MA, Steinbeck C. COCONUT online: Collection of Open Natural Products database. J Cheminform. 2021;13(1):2. Epub 20210110. doi: 10.1186/s13321-020-00478-9.
28. Schymanski EL, Kondic T, Neumann S, Thiessen PA, Zhang J, Bolton EE. Empowering large chemical knowledge bases for exposomics: PubChemLite meets MetFrag. J Cheminform. 2021;13(1):19. Epub 20210308. doi: 10.1186/s13321-021-00489-0.
29. Wishart DS, Guo A, Oler E, Wang F, Anjum A, Peters H, Dizon R, Sayeeda Z, Tian S, Lee BL, Berjanskii M, Mah R, Yamamoto M, Jovel J, Torres-Calzada C, Hiebert-Giesbrecht M, Lui VW, Varshavi D, Varshavi D, Allen D, Arndt D, Khetarpal N, Sivakumaran A, Harford K, Sanford S, Yee K, Cao X, Budinski Z, Liigand J, Zhang L, Zheng J, Mandal R, Karu N, Dambrova M, Schioth HB, Greiner R, Gautam V. HMDB 5.0: the Human Metabolome Database for 2022. Nucleic Acids Res. 2022;50(D1):D622-D31. doi: 10.1093/nar/gkab1062.
30. Mohammed Taha H, Aalizadeh R, Alygizakis N, Antignac JP, Arp HPH, Bade R, Baker N, Belova L, Bijlsma L, Bolton EE, Brack W, Celma A, Chen WL, Cheng T, Chirsir P, Cirka L, D'Agostino LA, Djoumbou Feunang Y, Dulio V, Fischer S, Gago-Ferrero P, Galani A, Geueke B, Glowacka N, Gluge J, Groh K, Grosse S, Haglund P, Hakkinen PJ, Hale SE, Hernandez F, Janssen EM, Jonkers T, Kiefer K, Kirchner M, Koschorreck J, Krauss M, Krier J, Lamoree MH, Letzel M, Letzel T, Li Q, Little J, Liu Y, Lunderberg DM, Martin JW, McEachran AD, McLean JA, Meier C, Meijer J, Menger F, Merino C, Muncke J, Muschket M, Neumann M, Neveu V, Ng K, Oberacher H, O'Brien J, Oswald P, Oswaldova M, Picache JA, Postigo C, Ramirez N, Reemtsma T, Renaud J, Rostkowski P, Rudel H, Salek RM, Samanipour S, Scheringer M, Schliebner I, Schulz W, Schulze T, Sengl M, Shoemaker BA, Sims K, Singer H, Singh RR, Sumarah M, Thiessen PA, Thomas KV, Torres S, Trier X, van Wezel AP, Vermeulen RCH, Vlaanderen JJ, von der Ohe PC, Wang Z, Williams AJ, Willighagen EL, Wishart DS, Zhang J, Thomaidis NS, Hollender J, Slobodnik J, Schymanski EL. The NORMAN Suspect List Exchange (NORMAN-SLE): facilitating European and worldwide collaboration on suspect screening in high resolution mass spectrometry. Environ Sci Eur. 2022;34(1):104. Epub 20221021. doi: 10.1186/s12302-022-00680-6.
31. Hirschfeld T. Instrumentation in the next decade. Science. 1985;230(4723):286-91. doi: 10.1126/science.230.4723.286.

32. Bremer PL, Wohlgemuth G, Fiehn O. The BinDiscover database: a biology-focused meta-analysis tool for 156,000 GC-TOF MS metabolome samples. J Cheminform. 2023;15(1):66. Epub 20230720. doi: 10.1186/s13321-023-00734-8.
33. Quinn RA, Nothias LF, Vining O, Meehan M, Esquenazi E, Dorrestein PC. Molecular Networking As a Drug Discovery, Drug Metabolism, and Precision Medicine Strategy. Trends Pharmacol Sci. 2017;38(2):143-54. Epub 20161111. doi: 10.1016/j.tips.2016.10.011.
34. For chemists, the AI revolution has yet to happen. Nature. 2023;617(7961):438. doi: 10.1038/d41586-023-01612-x.

DECODING THE MOLECULAR UNIVERSE
# 2023 WORKSHOP

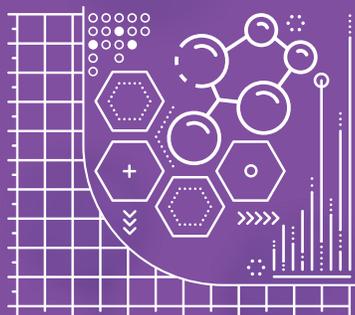

m/q
INITIATIVE
@PNNL

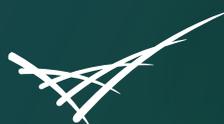

Pacific
Northwest
NATIONAL LABORATORY